# Speed of sound for three binary ($CH_4$ + $H_2$) mixtures from $p$ = (0.5 up to 20) MPa at $T$ = (273.16 to 375) K

Daniel Lozano-Martín[a], M. Carmen Martín[a], César R. Chamorro[a], Dirk Tuma[b], José Juan Segovia[a]

[a] BioEcoUVa, Research Institute on Bioeconomy, TERMOCAL Research Group, University of Valladolid, Escuela de Ingenierias Industriales, Paseo del Cauce 59, 47011 Valladolid, Spain

[b] BAM Bundesanstalt für Materialforschung und -prüfung, Berlin, D-12200, Germany.

## Abstract

Speed of sound is one of the thermodynamic properties that can be measured with least uncertainty and is of great interest in developing equations of state. Moreover, accurate models are needed by the $H_2$ industry to design the transport and storage stages of hydrogen blends in the natural gas network. This research aims to provide accurate data for ($CH_4$ + $H_2$) mixtures of nominal (5, 10, and 50) mol-% of hydrogen, in the $p$ = (0.5 up to 20) MPa pressure range and with temperatures $T$ = (273.16, 300, 325, 350, and 375) K. Using an acoustic spherical resonator, speed of sound was determined with an overall relative expanded ($k$ = 2) uncertainty of 220 parts in $10^6$ (0.022 %). Data were compared to reference equations of state for natural gas-like mixtures, such as AGA8-DC92 and GERG-2008. Average absolute deviations below 0.095% and percentage deviations between 0.029% and up to 0.30%, respectively, were obtained. Additionally, results were fitted to the acoustic virial equation of state and adiabatic coefficients, molar isochoric heat capacities and molar isobaric heat capacities as perfect-gas, together with second and third acoustic virial coefficients were estimated. Density second virial coefficients were also obtained.

## Keywords

Speed of sound; acoustic resonator; methane; hydrogen; heat capacities as perfect gas; virial coefficients

| *Nomenclature* | |
|---|---|
| $a$ | Inner radius of the cavity, m |
| $A_i$ | Coefficients of acoustic virial equation |
| $b$ | Outer radius of the cavity, m |
| $B(T)$ | Second virial coefficient, $cm^3 \cdot mol^{-1}$ |
| $C_p$ | Specific isobaric heat capacity, $J \cdot kg^{-1} \cdot K^{-1}$ |
| $C_{p,w}$ | Specific isobaric heat capacity of the wall material, $J \cdot kg^{-1} \cdot K^{-1}$ |
| $C_V$ | Specific isochoric heat capacity, $J \cdot kg^{-1} \cdot K^{-1}$ |
| $C_{p,m}$ | Molar isobaric heat capacity, $J \cdot mol^{-1} \cdot K^{-1}$ |
| $C_{V,m}$ | Molar isochoric heat capacity, $J \cdot mol^{-1} \cdot K^{-1}$ |
| $E$ | Young's modulus, Pa |
| $f_{0n}$ | Resonance frequency of acoustic $(0,n)$ mode, Hz |
| $g_{0n}$ | Resonance halfwidth of acoustic $(0,n)$ mode, Hz<br>Gravitational acceleration, $m \cdot s^{-2}$ |
| $F$ | Complex resonance frequency, Hz |
| $h$ | Thermal accommodation coefficient ($h = 1$ for mixtures of polyatomic gases) |
| $h_P$ | Planck Constant, J·s |
| $k$ | Coverage factor |
| $k_B$ | Boltzmann Constant, $J \cdot K^{-1}$ |
| $m$ | Mass, kg |
| $L$ | Duct length, m |
| $M$ | Molar Mass, kg/mol |
| N | Number of components of a mixture |
| $p$ | Pressure, MPa |
| $r_0$ | Duct radius, m |
| $r_{tr}$ | Radius of the transducer, m |

| *Greek symbols* | |
|---|---|
| $\alpha$ | Reduced Helmholtz free energy<br>Thermal expansion coefficient, $K^{-1}$ |
| $\beta_a$ | 2nd acoustic virial coefficient, $m^3 \cdot mol^{-1}$ |
| $\Delta$ | Frequency perturbation, Hz |
| $\gamma$ | Adiabatic coefficient |
| $\gamma_a$ | 3rd acoustic virial coefficient, $m^6 \cdot mol^{-2}$ |
| $\gamma_{eff}$ | Effective adiabatic coefficient ($\gamma_{eff} = 1$) |
| $\delta$ | Reduced density |
| $\eta$ | Shear viscosity, Pa·s |
| $\kappa$ | Thermal conductivity, $W \cdot m^{-1} \cdot K^{-1}$ |
| $\kappa_T$ | Isothermal compressibility, $Pa^{-1}$ |
| $\kappa_w$ | Thermal conductivity of the wall material, $W \cdot m^{-1} \cdot K^{-1}$ |
| $\nu_{0n}$ | Acoustic radial mode eigenvalue |
| $f$ | Drive frequency generated by the wave generator, Hz |
| $\nu_i$ | Molecular vibrational frequency, Hz |
| $\rho$ | Density, $kg \cdot m^{-3}$ |
| $\rho_n$ | Molar density, $mol \cdot m^{-3}$ |
| $\rho_w$ | Density of the wall material, $kg \cdot m^{-3}$ |
| $\sigma$ | Poisson's ratio |
| $\tau$ | Reduced temperature |
| $\tau_{vib}$ | Vibrational relaxation time, s |

| *Subscripts* | |
|---|---|
| 0 | Reference state |
| 0n | Acoustic radial mode index |

| | | | | |
|---|---|---|---|---|
| $R$ | Molar gas constant, $J \cdot mol^{-1} \cdot K^{-1}$ | | 1 | Component 1 of a binary mixture |
| $s$ | Standard deviation | | 2 | Component 2 of a binary mixture |
| $T$ | Temperature, K | | AGA8-DC92 | Calculated from AGA8-DC92 equation of state |
| $u$ | Standard uncertainty | | c | Critical parameter |
| $U$ | Expanded uncertainty | | EoS | Calculated from an equation of state |
| $V$ | Volume, $m^3$ | | exp | Experimental data |
| $V_h$ | Volume of the holes drilled in the transducer backplate, $m^3$ | | GERG-2008 | Calculated from GERG-2008 equation of state |
| $w$ | Speed of sound, $m \cdot s^{-1}$ | | r | Relative |
| $w_w$ | Speed of sound in the wall material, $m \cdot s^{-1}$ | | th | Thermal boundary layer |
| $Z$ | Compressibility factor | | sh | Shell |
| *Abbreviations* | | | | |
| BAM | Federal Institute for Materials Research and Testing | | *Superscripts* | |
| CEM | National Metrology Institute of Spain | | 0, *pg* | Ideal gas behavior |
| GUM | Guide to the Expression of Uncertainty in Measurement | | r | Residual behavior |

## 1. Introduction

Integrating hydrogen into the natural gas grid has been discussed in some recent projects [1-2] as a promising way of providing a sustainable and cost-effective energy source. Adding hydrogen to natural gas pipelines overcomes the important drawback related to the high costs of a dedicated hydrogen grid designed to transport and store hydrogen for end-users. Additional benefits of hydrogen injection include a significant reduction in carbon dioxide emissions of natural gas applications when hydrogen is produced from renewable sources or from steam reforming of fossil fuels with carbon capture and storage systems. Moreover, hydrogen can be used directly in electric vehicle fuel cells and stationary power units after extraction from the natural gas blend, thereby improving air quality due to the reduction in the sulfur dioxide, oxides of nitrogen and volatile organic compounds produced by traditional consumption of fossils fuels [3]. It is widely agreed that a 5 vol-% hydrogen blend does not require any modifications in natural gas line components, and the concentration can be increased up to (10 - 15) vol-% with minor only modifications [4-5]. As regards the critical parameters of gas burners, a 10 % hydrogen blend reduces the Wobbe index by about 3 % compared to a natural gas without hydrogen. As for the critical parameters of gas engines, the methane number of pure methane, biogas, biomethane or liquefied natural gas displays greater variability between these fuels than the effect of adding hydrogen. Furthermore, a typical 5 % increase in the laminar flame speed is obtained for a 10 % hydrogen blend [4].

All studies [2-3,6] recommend a case-by-case basis research before introducing hydrogen into natural gas pipelines, with an accurate estimation of the thermodynamic properties of natural gas and hydrogen blends being required for this purpose. The thermodynamic reference models for natural gas mixtures which include hydrogen are the GERG-2008 Equation of State (EoS) [7-8], an ISO standard [9], and the widely used AGA8-DC92 EoS [10]. The multiparametric GERG-2008 mixture model [8] is designed to predict volumetric and caloric properties in the gas, liquid or supercritical homogeneous region (and vapor-liquid equilibrium, VLE) of a mixture composed of the 21 major and minor components present in natural gas, and the full range of application covers the pressure range from (0 to 70) MPa and temperature range from (60 to 700) K. Binary interactions are introduced into these equations by composition-dependent mixture reducing functions for the density and temperature fitted to selected experimental data. When highly accurate binary data exist in the literature, a generalized or a specific departure function is regressed and added to the model. For the binary methane + hydrogen mixtures, a binary specific departure function was developed, and the binary interaction coefficients are fitted using density ($p,\rho,T$) and vapor-liquid equilibrium data. The density points cover the gas region with hydrogen molar fractions $x_{H_2}$ above 0.15 and temperatures above 270 K, and the liquid region with hydrogen molar fractions above 0.05 and temperatures as low as 130 K. Some recent studies have analyzed the performance of the AGA8-DC92 [10] and GERG-2008 [8] models, estimating the thermophysical properties of hydrogen + natural gas mixtures related to the study of density [11], and concerning the propagation of a decompression along a pipeline [12], an issue in which speed of sound plays a key role.

The aim of this work is to assess the performance of both GERG-2008 [8] and AGA8-DC92 EoS [10] for methane + hydrogen mixtures with a broader range of hydrogen content than the experimental data previously used to develop the two models. In order to meet this objective, speed of sound measurements were performed with one of the most accurate experimental setups, an acoustic spherical resonator, for three nominal molar compositions of (0.95 $CH_4$ + 0.05 $H_2$), (0.90 $CH_4$ + 0.10 $H_2$), and (0.50 $CH_4$ + 0.50 $H_2$) in the pressure range $p$ = (0.5 up to 20) MPa for temperatures $T$ = (273.16, 300, 325, 350, and 375) K. The uncertainty of the AGA8-DC92 model [10] with regard to speed of sound is stated to be 0.2 % for natural gas mixtures for pressures below 5 MPa and up to 0.8 % for higher pressures. The uncertainty of the GERG-2008 EoS [8] in terms of speed of sound for a binary mixture with a binary specific departure function is stated to be 0.1 % in the gas phase region. In addition, this research aims to expand the thermodynamic database, should any new correlations be carried out in future, since no experimental speed of sound data are found in the literature for methane + hydrogen mixtures [13]. Furthermore, heat capacities as perfect gas, acoustic and density virial coefficients derived from speed of sound measures are provided, which can be used to improve and develop thermodynamic models to facilitate the integration of hydrogen into the natural gas grid.

## 2. Materials and methods

### 2.1 Mixture preparation

Table 1 shows the composition and corresponding expanded ($k$ = 2) uncertainty of the three binary ($CH_4$ + $H_2$) mixtures studied in this work. These were prepared at the Federal Institute for Materials Research and Testing (BAM) in Germany using the gravimetric method, and the composition was validated by gas chromatography (GC) (ISO 6142-1 [14]). The relative deviations of the gravimetric composition given in Table 1 from the composition determined by GC analysis are within the stated expanded ($k$ = 2) uncertainty of the composition. The purity of the pure gases used is also indicated in Table 1, and more details concerning the mixture preparation procedure are reported elsewhere [15]. These were homogenized by thermal mixing and rolling during preparation, and were measured within six months of delivery. Mixtures were used without further purification.

**Table 1.** Mole fraction $x_i$ and expanded ($k$ = 2) uncertainty $U(x_i)$ of the binary ($CH_4$ + $H_2$) mixtures studied in this work. Purity, supplier, and critical temperature $T_c$ and pressure $p_c$ of the pure components used for the realization of the binary mixtures at BAM are also reported.

| Composition | (0.95 $CH_4$ + 0.05 $H_2$) | | (0.90 $CH_4$ + 0.10 $H_2$) | | (0.50 $CH_4$ + 0.50 $H_2$) | |
|---|---|---|---|---|---|---|
| | $x_i \cdot 10^2$ / mol/mol | $U(x_i) \cdot 10^2$ / mol/mol | $x_i \cdot 10^2$ / mol/mol | $U(x_i) \cdot 10^2$ / mol/mol | $x_i \cdot 10^2$ / mol/mol | $U(x_i) \cdot 10^2$ / mol/mol |
| Methane | 94.9914 | 0.0021 | 90.0034 | 0.0021 | 49.9678 | 0.0016 |
| Hydrogen (normal) | 5.0086 | 0.0018 | 9.9966 | 0.0028 | 50.0322 | 0.0128 |

| Pure Components | Supplier | Purity / mol/mol | $T_c$ / K$^{(*)}$ | $p_c$ / MPa$^{(*)}$ |
|---|---|---|---|---|
| Methane | Linde | ≥ 0.999995 | 190.564 | 4.5992 |
| Hydrogen (normal) | Linde | ≥ 0.999999 | 33.145 | 1.2964 |

$^{(*)}$ The critical parameters are computed from Refprop [16] using the reference EoS for methane [17] and hydrogen [18].

### 2.2 Equipment description

Speed of sound measurements $w(p,T,\bar{x})$ were performed with a spherical acoustic resonator of nominal internal radius $a$ = 40 mm manufactured at the Imperial College of London workshop using austenitic grade A321 stainless-steel. The experimental setup, including the acoustic cavity, transducers and thermostat device, is depicted in figure 1. Both the acoustic cavity and the transducer designs are based on those used by Trusler, Wakeham and Zarari [19-21] to measure pure methane and related mixtures. Further details concerning the setup may be found in [22-24], while only a brief description is given here.

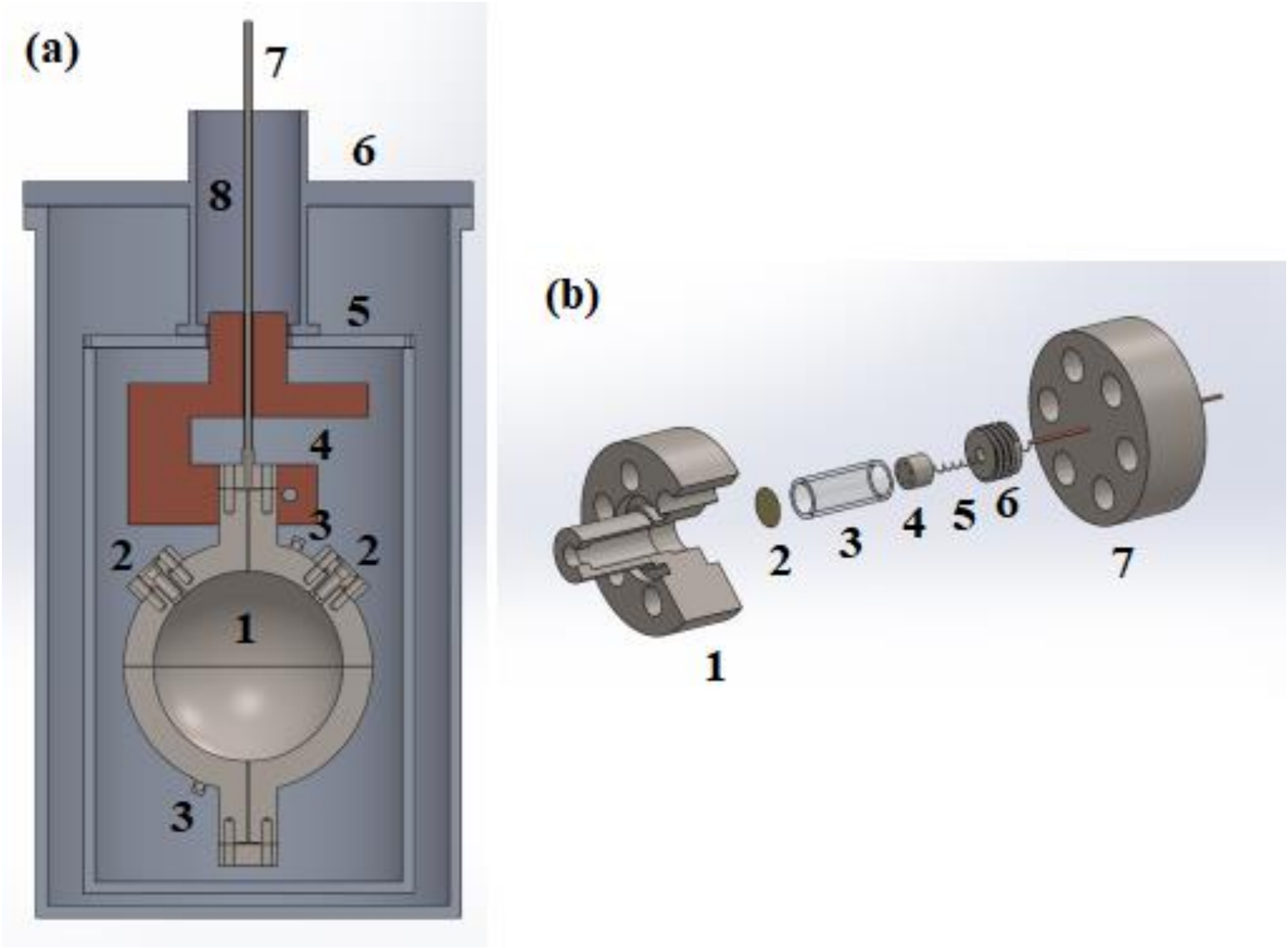

**Figure 1.** Schematic plots of the: (a) acoustic resonance cavity and thermostat setup - 1. spherical resonance cavity, 2. acoustic transducers, 3. standard platinum resistance thermometers (SPRTs), 4. copper block, 5. internal shell, 6. external shell, 7. gas inlet duct, 8. to vacuum; and (b) acoustic transducers - 1. transducer housing, 2. solid-dielectric diaphragm, 3. glass sleeve, 4. backplate, 5. spring of electrical contact, 6. screw, 7. cover plate.

Transducers are allocated forming an angle of 90º between them, with each being 45º from the north pole, in an attempt to prevent overlapping between the lowest radial acoustic mode (0,2) and the degenerate (0,3) mode closest to it [25]. There can be two types of geometrical imperfections in the acoustic cavity: those caused by machining errors, which may lead to spheroidal distortions, hemispheres with unequal radii or misalignment [26], and those arising from annular slits surrounding the transducer ports [27] or gaps in the equatorial junction of the hemispheres as a result of the welding having failed to penetrate fully [28]. In both cases, standard machining should lead to imperfections that are no worse than 1 part in $10^4$ of the internal radius, which is reflected in speed of sound determination with an error better than 1 part in $10^6$, according to a perfect sphere. These effects are, therefore, negligible.

Both source and detector acoustic transducers are equal solid dielectric capacitance type transducers which are made ad-hoc for speed of sound determination and are located nearly flush with the internal surface of the resonance cavity. This assembly is a high acoustic impedance, low output power and wide frequency bandwidth device with mechanical resonance frequencies of around 40 kHz. The upper operating temperature is restricted to below 475 K due to the degradation of the dielectric membrane.

The source transducer is excited by the ac sinusoidal signal without bias voltage sent out by a wave synthesizer HP3225B. In the absence of dc bias, resonance frequency is detected by the receiver at twice the driven frequency of the transmitter, avoiding the undesirable crosstalk effect between source and detector transducers. The receiver transducer is connected to the amplifier by a triaxial cable to compensate the division of the signal by the high capacitance of the long cables compared to the small capacitance load of the transducer (below 100 pF). The output signal from the amplifier is measured by a dual-phase Lock-In SR850 DSP amplifier detector operating at differential voltage mode with its reference to the second harmonic of the wave generator. The Lock-In processes the input signal and decomposes in the in-phase $u$ and quadrature $v$ components.

Pressure is controlled by two piezoelectric quartz transducers, a Digiquartz 2003A-101 for pressures below 2 MPa, and a Digiquartz 43KR-101 for pressures from (2 up to 20) MPa, calibrated against a dead weight pneumatic pressure balance. The expanded ($k$ = 2) uncertainty in pressure is between (2.4 to 17)·$10^{-4}$ MPa. Temperature is controlled by a thermostatic setup composed of an internal shell designed to prevent heat losses by radiation, an external shell which is made to provide a vacuum and avoid heat transfer by convection around the resonance cavity, and a copper block from which the cavity is suspended in order to control temperature only by heat conduction. The whole device is immersed in a Dewar with ethanol cooled so as to ensure thermal stabilization in the order of a few mK. Temperature is measured by two SPRT Rosemount 162D of 25.5 Ω calibrated on ITS-90 applying the procedure described in [29-30] and located in the northern and southern hemispheres of the acoustic cavity. Expanded ($k$ = 2) uncertainty in temperature is below $4.6 \cdot 10^{-3}$ K.

**2.3 Measurement procedure**

Speed of sound $w(p,T,\bar{x})$ is obtained at each state point from the resonance frequency $f_{ln}$ of the first five pure radial (non-degenerate) acoustic modes, labeled as ($l$ = 0,$n$) = (0,2), (0,3), (0,4), (0,5) and (0,6). Resonance frequency is determined by measuring the in-phase $u$ and quadrature $v$ signals at 11 equally spaced drive frequencies. The 22 values of the complex signal $u$ + i$v$ are fitted by a non-linear regression algorithm to a Lorentzian shape function with linear background level [31-32]:

$$u + iv = \frac{A^*}{\left(F^2 - f^2\right)} + B^* + C^* f \tag{1}$$

where $A^*$ is a complex fitted parameter proportional to the amplitude of the sound wave produced by the source detector and the sensitivity of the detector transducer, $B^*$ and $C^*$ are complex parameters, $F = f_{0n} + \mathrm{i}g_{0n}$, $f_{0n}$ is the resonance frequency, and $g_{0n}$ is the resonance halfwidth of (0,$n$) mode. The fit is performed with a constant background level $C^* = 0$ and a linear level $C^* \neq 0$, taking the results with the least regression error as the experimental resonance frequency and halfwidth.

The working equation that relates $w(p,T)$ to $f_{0n}$ is obtained applying the ideal case radial boundary condition of zero acoustic admittance and perfect geometry to the solution of the homogeneous wave equation:

$$w(p,T) = \frac{2\pi a(p,T)}{\nu_{0n}}(f_{0n} - \Delta f) \tag{2}$$

where $a$ is the internal radius of the resonance cavity, $\Delta f$ is the sum of all the frequency perturbations due to the different effects that provide a non-zero acoustic admittance and imperfect geometry of the resonance cavity, and $v_{0n}$ is the zero of spherical Bessel first derivative for the $n^{th}$ mode of order $l = 0$. The internal radius of the resonance cavity $a(p,T)$ as a function of pressure and temperature was obtained in a previous calibration [33-34] by speed of sound measurements in argon, using the same setup as described in this work and the estimated value of the speed of sound given by the reference EoS of argon [35]. The regression coefficients of the quadratic polynomial dependence on pressure of $a(p,T)$ for the isotherms studied in this research and the estimated expanded ($k$ = 2) relative uncertainty of the radius $U_r(a)$ can be consulted in table 4 of [34]. The main contribution to radius uncertainty is due to the uncertainty of the equation of state, which comes to 0.02 % in the speed of sound.

Frequency corrections $\Delta f$ and contributions to halfwidths $g$ consider the effects of the thermal boundary layer, coupling of the fluid and shell motion, perturbation of the ducts, perturbation of the transducers, classical dissipation in the fluid bulk, and molecular vibrational relaxation phenomena. The list of equations used to compute these effects is summarized in table 2 and is based on the theory developed in [28,36].

**Table 2.** Expressions for frequency corrections $\Delta f$ and halfwidths $g$ calculated by the acoustic model for a spherical cavity.

| Quantity | Relationship | |
|---|---|---|
| Thermal Boundary Layer | $\frac{\Delta f_{\mathrm{th}}}{f} = -\frac{\gamma-1}{2a}\delta_{\mathrm{th}} + \frac{\gamma-1}{a} l_{\mathrm{th}} + \frac{\gamma-1}{2a}\delta_{\mathrm{th,w}}\frac{\kappa}{\kappa_{\mathrm{w}}}$ | |
| | $\frac{g_{\mathrm{th}}}{f} = \frac{\gamma-1}{2a}\delta_{\mathrm{th}} + \frac{\gamma-1}{2a}\delta_{\mathrm{th,w}}\frac{\kappa}{\kappa_{\mathrm{w}}} - \frac{1}{2}(\gamma-1)(2\gamma-1)\left(\frac{\delta_{\mathrm{th}}}{a}\right)^2$ | |
| | Thermal Penetration Length / m | $\delta_{\mathrm{th}} = \left(\frac{\kappa}{\pi\rho C_{\mathrm{p}} f}\right)^{1/2}$ |
| | Thermal Accommodation Length / m | $l_{\mathrm{th}} = \frac{\kappa}{p}\left(\frac{\pi M T}{2R}\right)^{1/2}\frac{2-h}{h}\frac{1}{C_{\mathrm{V}} M / R + 1/2}$ |
| | Thermal Penetration Length for Wall Material / m | $\delta_{\mathrm{th,w}} = \left( \frac{\kappa_{\mathrm{w}}}{\pi \rho_{\mathrm{w}} C_{\mathrm{p,w}} f} \right)^{1/2}$ |
| Bulk Dissipation | $\frac{g_{\mathrm{cl}}}{f} = f^2 \frac{\pi^2}{w^2} \left[ \frac{4}{3} \delta_{\mathrm{s}}^2 + (\gamma - 1) \delta_{\mathrm{th}}^2 \right]$ | |
| | Viscous Penetration Length / m | $\delta_{\mathrm{s}} = \left( \frac{\eta}{\rho \pi f} \right)^{1/2}$ |
| Coupling of Fluid and Shell Motion | $\frac{\Delta f_{\mathrm{sh}}}{f} = -\frac{\rho w^2}{\rho_{\mathrm{w}} w_{\mathrm{w}}^2} q \frac{(1 + AB - qB^2)\tan(B - A) - (B - A) - qAB^2}{\left[(qA^2 - 1)(qB^2 - 1) + AB\right]\tan(B - A) - (1 + qAB)(B - A)}$ <br> $q = \frac{1 - \sigma}{2(1 - 2\sigma)}$ <br> $A = \frac{2\pi f a}{w_{\mathrm{w}}}$ <br> $B = \frac{2\pi f b}{w_{\mathrm{w}}}$ <br> $w_{\mathrm{w}} = \left( \frac{1 - \sigma}{(1 - 2\sigma)(1 + \sigma)} \frac{E}{\rho_{\mathrm{w}}} \right)^{1/2}$ | |
| Inlet Gas Correction | $\frac{\Delta f_0 + ig}{f} = i \frac{r_0^2}{4a^2 \nu_{0\mathrm{n}}} y_0$ | |
| | Acoustic Admittance of the Opening | $y_0 = i \tan(k_{\mathrm{KH}} L)$ |
| | Kirchhoff-Helmholtz Propagation Constant / m$^{-1}$ | $k_{\mathrm{KH}} = \frac{2\pi f}{w} + (1 - i)\left( \frac{\pi f}{w r_0} \left[ \delta_{\mathrm{s}} + (\gamma - 1)\delta_{\mathrm{th}} \right] \right)$ |
| Transducer Correction | $\frac{\Delta f_{\mathrm{tr}}}{f} = -\frac{\rho w^2 r_{\mathrm{tr}}^2}{2a^3} \mathrm{X}_{\mathrm{m}}$ | |
| | Acoustic Admittance of the Transducer | $y_{\mathrm{tr}} = i\omega \rho w \mathrm{X}_{\mathrm{M}}$ |
| | Compliance of Transducer / m·Pa$^{-1}$ | $\mathrm{X}_{\mathrm{m}} = \frac{V_{\mathrm{h}} \kappa_{\mathrm{T}}}{\gamma_{\mathrm{eff}} \pi r_{\mathrm{tr}}^2}$ |
| Vibrational Correction | $\frac{\Delta f_{\mathrm{vib}}}{f} = \frac{1}{2}(\gamma - 1)\Delta(2\pi f \tau_{\mathrm{vib}})^2 \left( 1 - \frac{\Delta(1 + 3\gamma)}{4} \right)$ | |
| | Vibrational Relaxation Time / s | $\tau_{\mathrm{vib}} = \frac{\Delta g}{\Delta(\gamma - 1)\pi f^2} = \frac{g - (g_{\mathrm{th}} + g_{\mathrm{cl}} + g_0)}{\Delta(\gamma - 1)\pi f^2}$ |
| | Vibrational Contribution to Isobaric Heat Capacity of the Mixture | $\Delta = \sum_{\mathrm{k}} x_{\mathrm{K}} \frac{C_{\mathrm{vib,k}}}{C_{\mathrm{p}} M}$ |
| | Vibrational Heat Capacity (Plank-Einstein function) / J·mol$^{-1}$·K$^{-1}$ | $C_{\mathrm{vib,k}} = R \sum_{\mathrm{i}} \frac{z_{\mathrm{i}}^2 e^{z_{\mathrm{i}}}}{(e^{z_{\mathrm{i}}} - 1)^2}$ |
| | $z_{\mathrm{i}} = \frac{\vartheta_{\mathrm{i}}}{T} = \frac{h_{\mathrm{P}} \nu_{\mathrm{i}} / k_{\mathrm{B}}}{T}$ | |

Perturbations in frequency take negative values for the acoustic modes (0,2), (0,3), (0,4) below the radial-symmetric mechanical resonance frequency of the cavity and positive values for the (0,6) above mode. Perturbation for the (0,5) mode is negative in the case of the (5 and 10) mol-% hydrogen mixture and positive for the other gas sample. Corrections range from (−850 to −50) parts in $10^6$ and from (1.5 to 520) parts in $10^6$ and increase smoothly as the hydrogen content and temperature rise and decrease towards lower pressures. At high pressures, perturbations increase with frequency, whereas at low pressures the behavior is the opposite. The thermodynamic and transport properties of the gas mixtures required for the calculations were estimated from the GERG-2008 mixture model reference [8] and the respective transport properties model references using Refprop 9.1 [16]. The transport and elastic properties of the stainless-steel wall material were estimated from the literature [37-39] assuming that grade A321 steel (cavity material) behaves the same as A304 stainless steel (almost similar composition).

The thermal boundary layer perturbation is the most significant at low pressures. One factor required to compute this correction is the thermal accommodation coefficient $h$, which depends on the wall and the gas. This coefficient has only been determined for very pure gases in contact with some specific surfaces, which is not our case [40]. However, an accurate value is not needed since the high-pressure results obtained in this work are not sensitive to reasonable values of $h$. Thus, $h$ is assumed to be equal to 1.

Coupling of fluid and shell motion is the most important frequency correction at high pressures. The expression of $\Delta f_{sh}/f$ given in table 2 diverges near the mechanical resonance of the cavity where acoustic modes suffer from high energy absorption and major perturbation to the resonance frequency. The lowest radial mechanical resonance of the shell is referred to as the breathing mode of the cavity $f_{br}$ and is estimated to be around 27 kHz for our cavity from the approximated expression:

$$f_{\text{br}} = \frac{w_{\text{w}}}{2\pi a}\left(\frac{2\left[\left(\frac{b}{a}\right)^3 - 1\right]}{\left[\frac{b}{a} - 1\right]\left[1 + 2\left(\frac{b}{a}\right)^3\right]}\right)^{1/2} \tag{3}$$

where $w_{\text{w}}$ is the longitudinal speed of sound in the wall material, and $a$ and $b$ are the inner and outer radius of the shell. The value of $f_{\text{br}}$ is the lowest frequency for which the denominator of $\Delta f_{\text{sh}}/f$ vanishes. Acoustic modes close to this value are thus omitted from the speed of sound calculation.

Acoustic measurements with frequencies above 40 kHz were not performed in order to prevent major perturbations to the frequency and halfwidth of the resonance peaks by the mechanical resonance of the transducers. Thus, the (0,6) mode for the mixture (0.50 $CH_4$ + 0.50 $H_2$) was omitted.

Moreover, speed of sound measurements in certain gases, such as methane, require molecular vibrational correction [41]. We assume that all the molecules relax in unison with a single overall relaxation time $\tau_{\text{vib}}$, which is determined from the excess halfwidths $\Delta g$ defined as the experimental halfwidth minus the contributions from thermal boundary perturbation $g_{\text{th}}$, classical viscothermal dissipation in the fluid bulk $g_{\text{cl}}$, and energy absorption in the ducts $g_0$:

$$\Delta g = g_{0\text{n}} - \left(g_{\text{th}} + g_{\text{cl}} + g_0\right) \tag{4}$$

Estimation of the molar vibrational heat capacity $C_{\text{vib}}$ is deduced from Planck-Einstein functions with vibrational frequencies determined by spectroscopy techniques for methane and hydrogen [42]. As figure 2 shows at $T$ = 273.16 K, the vibrational effect becomes significant at the lowest pressures, with $\tau_{\text{vib}}$ values ranging from (0.015 up to 0.3)$\cdot 10^{-6}$ s. Analogous results are obtained for the other isotherms studied in this work.

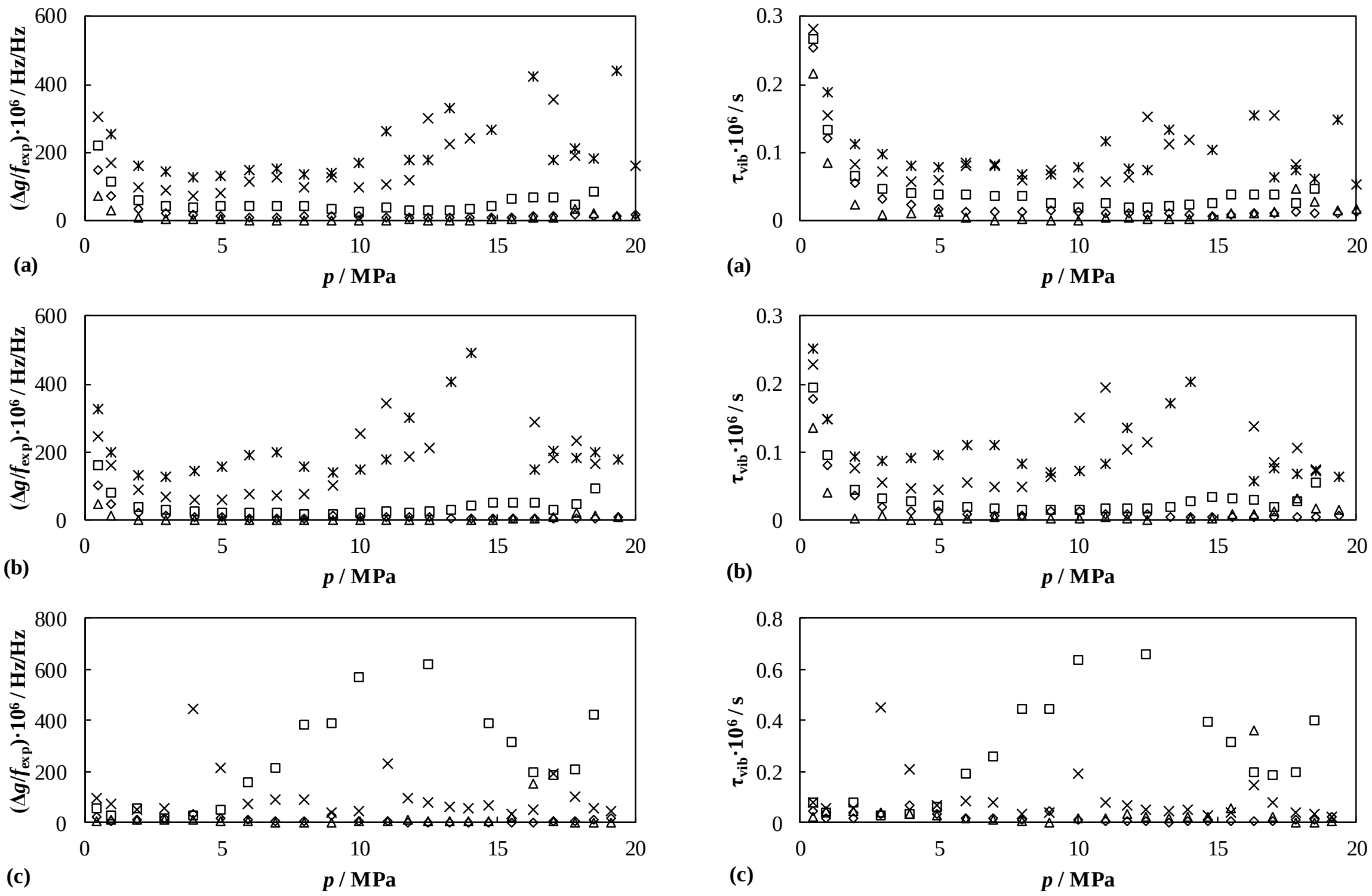

**Figure 2.** Relative excess halfwidths ($\Delta g/f$) (left figures), and relaxation times $\tau_{\mathrm{vib}}$ derived from $\Delta g$ (right figures) as a function of pressure due to vibrational relaxation at $T$ = 273.16 K for: (a) binary (0.95 $CH_4$ + 0.05 $H_2$), (b) (0.90 $CH_4$ + 0.10 $H_2$), and (c) (0.50 $CH_4$ + 0.50 $H_2$) mixtures, and for radial modes: △ (0,2), ◇ (0,3), □ (0,4), × (0,5), ✳ (0,6).

Speed of sound results at each temperature and pressure are obtained as the average of the individual values from equation (2) for each radial acoustic mode, although not all the measured modes have been used. As figure 2 shows, relative excess halfwidths $\Delta g/f$ of the (0,5) and (0,6) modes for (0.05 $CH_4$ + 0.95 $H_2$) and (0.10 $CH_4$ + 0.90 $H_2$) mixtures and the (0,4) mode for the (0.50 $CH_4$ + 0.50 $H_2$) mixture are significantly higher than the $\Delta g/f$ of the other modes, being some four to five times greater. This indicates that the acoustic model stated in table 2 to treat the experimental data is not enough to describe the behavior of these resonances. As a result, they are excluded from the final calculation of $w(p,T)$. The large $\Delta g/f$ of the discarded modes can be explained because they are close to the mechanical breathing mode of the shell, and the model applied to couple the fluid and shell motion does not take into account geometry modifications of perfect sphericity.

In addition, the increase in the acoustic resonance halfwidth when reducing the pressure due to the effects of the thermal boundary layer and vibrational relaxation means that the uncertainty involved in determining the resonance frequency is greater because a greater error emerges when fitting the signal to equation (1). After some tests, we decided not to perform measures below 0.5 MPa.

### 2.4 Stability of the mixture

A test was run to check the stability of the gas sample during the measurements taken in this work (results are depicted in figure 3). One possible source of systematic errors when determining intensive speed of sound arises from the possible different adsorption in the cavity wall of one of the components of the mixture compared to the other, involving a change in the molar mass of the mixture. Continuous recording of the resonance frequency of acoustic (0,3) mode at the sample gas bottle pressure and the lowest isotherm ($T$ = 273.16 K) for the (0.90 $CH_4$ + 0.10 $H_2$) mixture were therefore performed over one week, the time required to fully determine each isotherm. The maximum difference in frequency after eight days was 0.38 Hz, which corresponds to a 28 part in $10^6$ change and half the expanded ($k$ = 2) relative uncertainty contribution of the gas composition to the speed of sound uncertainty, as reported in table 4. This effect is therefore discarded. It should also be noted that a lower effect of the adsorption phenomena is expected at higher temperatures or lower hydrogen content.

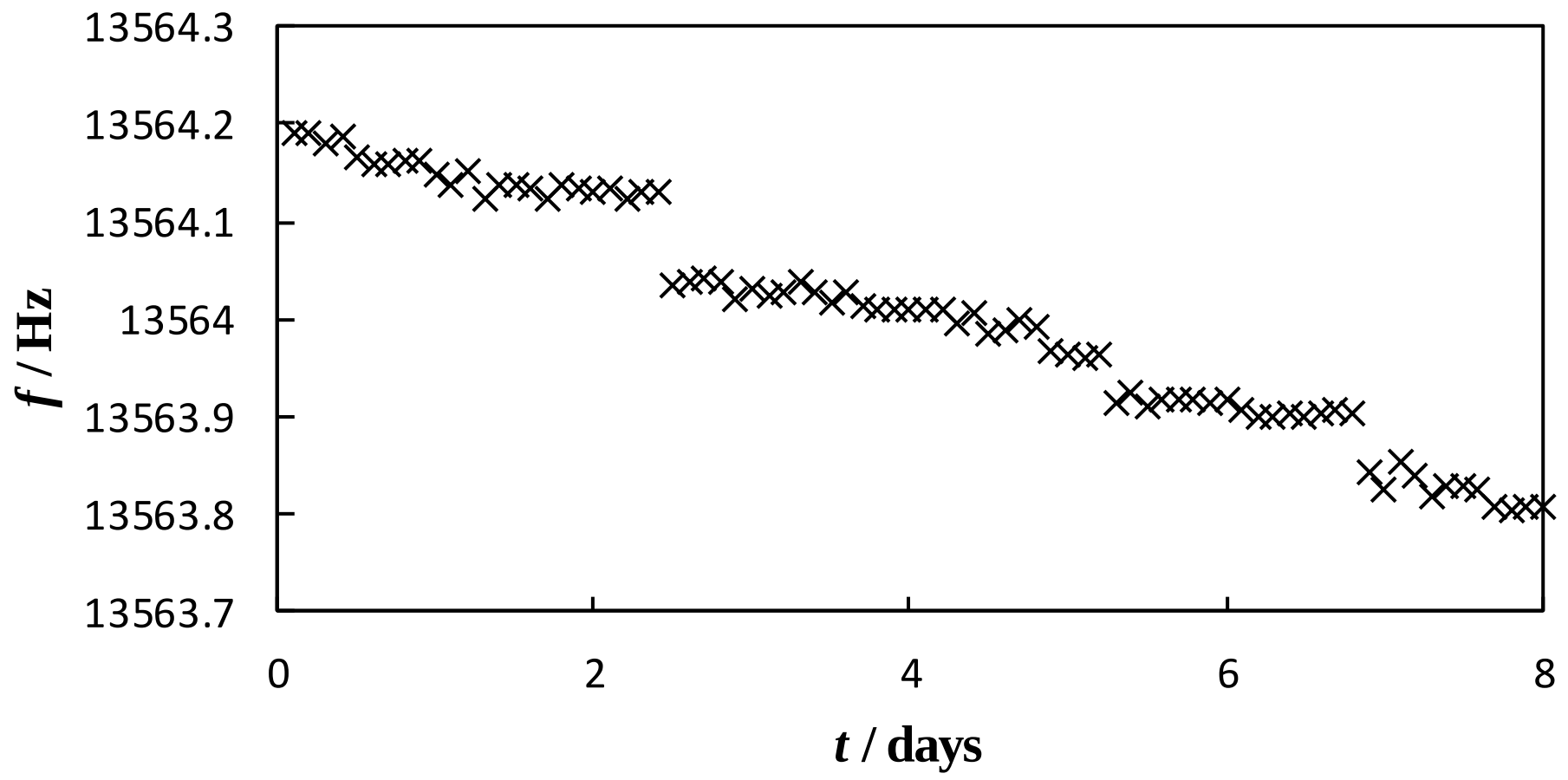

**Figure 3.** Frequency measurements as function of time (in days) performed for the assessment of the stability of the mixtures studied. They correspond to the acoustic (0,3) mode at $p$ ~ 7 MPa and $T$ = 273.16 K for the binary (0.90 $CH_4$ + 0.10 $H_2$) mixture.

## 3. Derived properties

Substituting the virial equation of state:

$$\frac{p}{\rho_{\mathrm{n}} RT} = 1 + B\rho_{\mathrm{n}} + C\rho_{\mathrm{n}}^2 + ... \tag{5}$$

into the equation that relates speed of sound with the thermodynamic state:

$$w(\rho_{\mathrm{n}}, T)^2 = \frac{RT}{M}\left[ Z + \rho_{\mathrm{n}}\left(\frac{\partial Z}{\partial \rho_{\mathrm{n}}}\right)_T + \frac{R}{C_{\mathrm{V,m}}}\left[ Z + T\left(\frac{\partial Z}{\partial T}\right)_{\rho_{\mathrm{n}}}\right]^2\right] \tag{6}$$

gives the expansion series in powers of $\rho_{\mathrm{n}}$:

$$w(\rho_{\mathrm{n}}, T)^2 = \frac{RT\gamma^{\mathrm{pg}}}{M}\left(1 + \beta_{\mathrm{a}}\rho_{\mathrm{n}} + \gamma_{\mathrm{a}}\rho_{\mathrm{n}}^2 + ...\right) \tag{7}$$

where $B(T)$, $C(T)$, …, are the density virial coefficients and $\beta_{\mathrm{a}}(T)$, $\gamma_{\mathrm{a}}(T)$, …, are the acoustic virial coefficients. Squared speed of sound data determined after frequency correction and data reduction are fitted to the acoustic virial equation expanded in powers of $p$:

$$w(p,T)^2 = A_0(T) + A_1(T)p + A_2(T)p^2 + ... \tag{8}$$

from where it is derived:

$$A_0 = \frac{RT\gamma^{\text{pg}}}{M} \tag{9}$$

$$\beta_a = A_1 \frac{M}{\gamma^{\text{pg}}} \tag{10}$$

$$\gamma_\text{a} = A_2 \frac{RTM}{\gamma^{\text{pg}}} + B\beta_\text{a} \tag{11}$$

where the superscript "*pg*" indicates perfect-gas. The experimental ideal gas heat capacities of the *N*-component mixture $C_{\text{p,mix}}^{\text{pg}}$ are compared to those of the reference AGA8-DC92 [10] and GERG-2008 EoS [8]:

$$\frac{C_{\text{p,mix}}^{\text{pg}}}{R} = \sum_{\text{i}}^{\text{N}} x_\text{i} \frac{\left(C_{\text{p,m}}^{\text{pg}}\right)_\text{i}}{R} \tag{12}$$

$$\frac{\left(C_{\text{p,m}}^{\text{pg}}\right)_\text{i}}{R} = B_\text{i} + C_\text{i}\left[\frac{D_\text{i}/T}{\sinh\left(D_\text{i}/T\right)}\right]^2 + E_\text{i}\left[\frac{F_\text{i}/T}{\cosh\left(F_\text{i}/T\right)}\right]^2 + G_\text{i}\left[\frac{H_\text{i}/T}{\sinh\left(H_\text{i}/T\right)}\right]^2 + I_\text{i}\left[\frac{J_\text{i}/T}{\cosh\left(J_\text{i}/T\right)}\right]^2$$

(13)

where the regression constants $B_\text{i}$ to $J_\text{i}$ are obtained for $i$ = $CH_4$ by fitting the spectroscopy data of McDowell and Kruse [43], and for $i$ = $H_2$ by fitting the data of Schäfer and Auer [44].

The polynomial order of the acoustic virial equation is increased according to the criteria that: the root mean square (RMS) of the residuals remain within experimental uncertainty; the *p*-value test of the analysis of variance (ANOVA) statistical table shows that all the regression parameters are significant; and no evidence of a systematic trend or pattern is found in the residuals. Consequently, these are randomly distributed as depicted in figure 4 for the binary (0.95 $CH_4$ + 0.05 $H_2$) mixture. Similar residual plots are obtained for the other two mixtures studied in this work, with RMS values of residuals that are below 39 parts in $10^6$ and well within experimental uncertainty. The regressed coefficients to equation (8) are given in table 3 together with their uncertainty estimated by the Monte Carlo method [45]. As the temperature is increased, the pressure dependence of the speed of sound is less sharp. For this reason, a lower polynomial order is required to fit the experimental data to equation (8). It is concluded that for the (0.95 $CH_4$ + 0.05 $H_2$) mixture, a sixth order is required at $T$ = 273.16 K, a fifth order at $T$ = (300 and 325) K, a fourth order at $T$ = 350 K, and a third order at $T$ = 375 K. For the (0.90 $CH_4$ + 0.10 $H_2$) mixture, a fifth order is necessary at $T$ = (273.16 and 300) K, a fourth order at $T$ = (325 and 350) K, and a third order at $T$ = 375 K. Finally, for the (0.50 $CH_4$ + 0.50 $H_2$) mixture, a fourth order is required for all the isotherms.

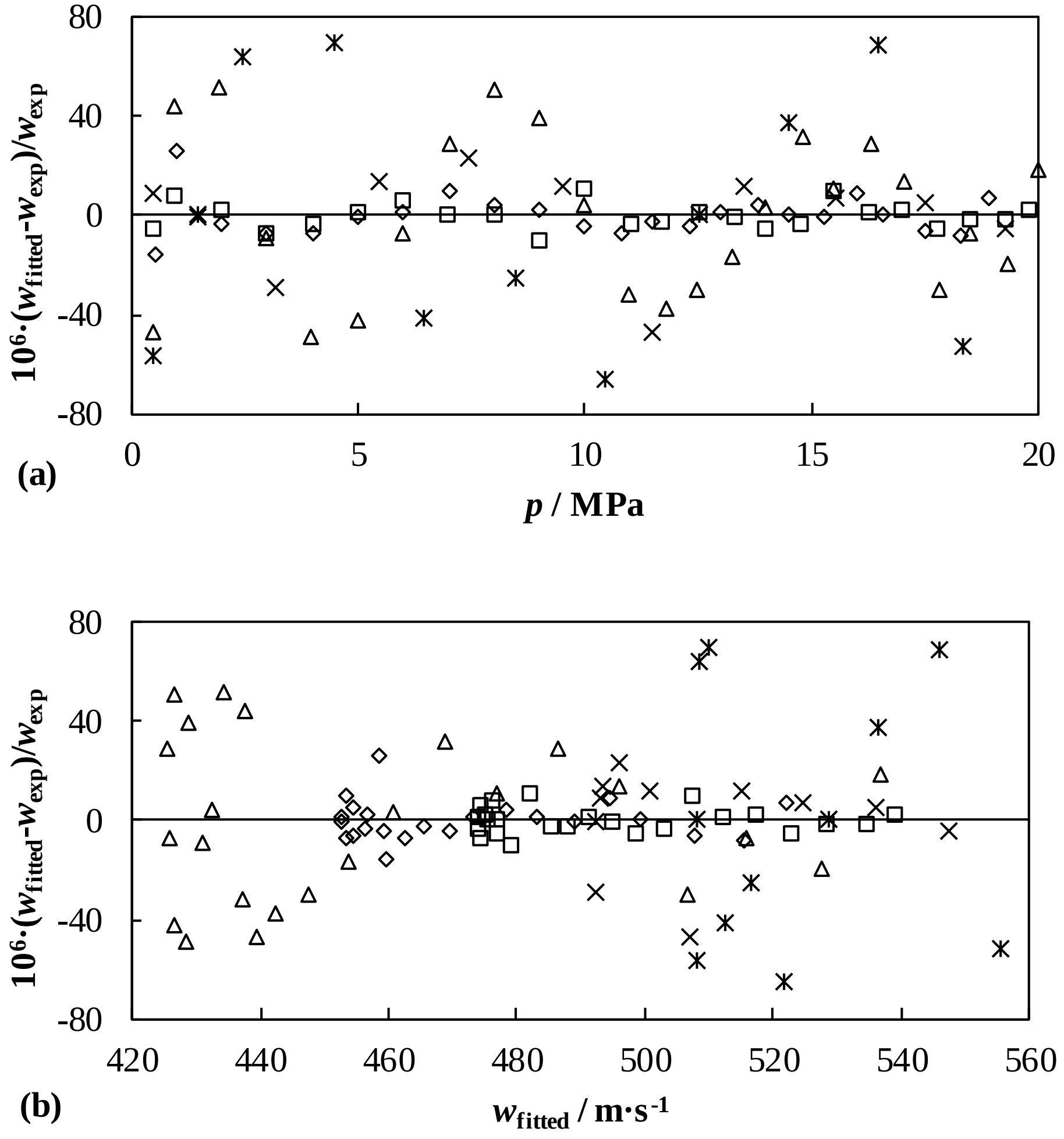

**Figure 4.** Residual plots $\Delta w = (w_{\text{fitted}} - w_{\text{exp}})/w_{\text{exp}}$ (a): as a function of the pressure (the independent variable), and (b): as a function of the fitted speed of sound from the values regressed to equation

(8), for the mixture (0.95 $CH_4$ + 0.05 $H_2$) at temperatures $T$ = △ 273.16 K, ◇ 300 K, □ 325 K, × 350 K, ✳ 375 K.

**Table 3.** Fitting parameters $A_i(T)$ of the square speed of sound to equation (8) and their corresponding expanded ($k$ = 2) uncertainties determined by the Monte Carlo method. The root mean square of the residuals of the fitting (RMS) is included on the last column.

| $T$ / K | $A_0(T)$ / $m^2 \cdot s^{-2}$ | $A_1(T)$ / $m^2 \cdot s^{-2} \cdot Pa^{-1}$ | $A_2(T)$ / $m^2 \cdot s^{-2} \cdot Pa^{-2}$ | $A_3(T)$ / $m^2 \cdot s^{-2} \cdot Pa^{-3}$ | $A_4(T)$ / $m^2 \cdot s^{-2} \cdot Pa^{-4}$ | $A_5(T)$ / $m^2 \cdot s^{-2} \cdot Pa^{-5}$ | $A_6(T)$ / $m^2 \cdot s^{-2} \cdot Pa^{-6}$ | RMS / ppm |
|---|---|---|---|---|---|---|---|---|
| | | | | (0.95 $CH_4$ + 0.05 $H_2$) | | | | |
| 273.16 | 194963 ± 29 | $(-372.7 \pm 4.9) \cdot 10^{-5}$ | $(23.2 \pm 2.4) \cdot 10^{-11}$ | $(-10.9 \pm 4.8) \cdot 10^{-18}$ | $(25.5 \pm 4.7) \cdot 10^{-25}$ | $(-9.4 \pm 2.1) \cdot 10^{-32}$ | $(8.6 \pm 3.7) \cdot 10^{-40}$ | 32 |
| 300.00 | 212402 ± 25 | $(-244.1 \pm 2.9) \cdot 10^{-5}$ | $(159.5 \pm 9.6) \cdot 10^{-12}$ | $(4.7 \pm 1.3) \cdot 10^{-18}$ | $(39.4 \pm 7.6) \cdot 10^{-26}$ | $(-13.7 \pm 1.6) \cdot 10^{-33}$ | - | 8 |
| 325.00 | 228104 ± 25 | $(-153.2 \pm 2.8) \cdot 10^{-5}$ | $(139.0 \pm 8.7) \cdot 10^{-12}$ | $(5.5 \pm 1.1) \cdot 10^{-18}$ | $(5.1 \pm 6.1) \cdot 10^{-26}$ | $(-4.0 \pm 1.2) \cdot 10^{-33}$ | - | 5 |
| 350.00 | 243462 ± 41 | $(-81.5 \pm 3.5) \cdot 10^{-5}$ | $(122.9 \pm 7.8) \cdot 10^{-12}$ | $(53.9 \pm 6.2) \cdot 10^{-19}$ | $(-9.1 \pm 1.6) \cdot 10^{-26}$ | - | - | 20 |
| 375.00 | 258514 ± 39 | $(-35.7 \pm 2.1) \cdot 10^{-5}$ | $(147.8 \pm 2.7) \cdot 10^{-12}$ | $(113.6 \pm 9.7) \cdot 10^{-20}$ | - | - | - | 50 |
| | | | | (0.90 $CH_4$ + 0.10 $H_2$) | | | | |
| 273.16 | 204726 ± 24 | $(-310.2 \pm 3.1) \cdot 10^{-5}$ | $(16.1 \pm 1.1) \cdot 10^{-11}$ | $(5.8 \pm 1.5) \cdot 10^{-18}$ | $(76.2 \pm 9.1) \cdot 10^{-26}$ | $(-26.1 \pm 1.9) \cdot 10^{-33}$ | - | 32 |
| 300.00 | 223173 ± 24 | $(-197.9 \pm 2.7) \cdot 10^{-5}$ | $(150.4 \pm 8.5) \cdot 10^{-12}$ | $(6.8 \pm 1.1) \cdot 10^{-18}$ | $(10.0 \pm 6.1) \cdot 10^{-26}$ | $(-6.1 \pm 1.2) \cdot 10^{-33}$ | - | 25 |
| 325.00 | 239590 ± 22 | $(-107.9 \pm 1.7) \cdot 10^{-5}$ | $(123.7 \pm 3.6) \cdot 10^{-12}$ | $(73.1 \pm 2.8) \cdot 10^{-19}$ | $(-124.3 \pm 7.2) \cdot 10^{-27}$ | - | - | 29 |
| 350.00 | 255745 ± 43 | $(-43.9 \pm 3.6) \cdot 10^{-5}$ | $(127.6 \pm 7.9) \cdot 10^{-12}$ | $(43.0 \pm 6.2) \cdot 10^{-19}$ | $(-7.4 \pm 1.6) \cdot 10^{-26}$ | - | - | 19 |
| 375.00 | 271518 ± 43 | $(2.0 \pm 2.1) \cdot 10^{-5}$ | $(144.9 \pm 2.5) \cdot 10^{-12}$ | $(82.7 \pm 8.5) \cdot 10^{-19}$ | - | - | - | 70 |
| | | | | (0.50 $CH_4$ + 0.50 $H_2$) | | | | |
| 273.16 | 339922 ± 25 | $(160.8 \pm 2.0) \cdot 10^{-5}$ | $(151.0 \pm 4.1) \cdot 10^{-12}$ | $(47.2 \pm 3.1) \cdot 10^{-19}$ | $(-110.5 \pm 7.6) \cdot 10^{-27}$ | - | - | 51 |
| 300.00 | 370796 ± 52 | $(247.5 \pm 4.7) \cdot 10^{-5}$ | $(11.8 \pm 1.1) \cdot 10^{-11}$ | $(39.8 \pm 8.3) \cdot 10^{-19}$ | $(-9.7 \pm 2.1) \cdot 10^{-26}$ | - | - | 69 |
| 325.00 | 398504 ± 63 | $(322.4 \pm 5.3) \cdot 10^{-5}$ | $(6.9 \pm 1.2) \cdot 10^{-11}$ | $(54.3 \pm 9.2) \cdot 10^{-19}$ | $(-13.1 \pm 2.4) \cdot 10^{-26}$ | - | - | 61 |

Second density virial coefficients $B(T)$ were derived from the perfect-gas heat capacities $C_{\mathrm{p,m}}^{\mathrm{pg}}$ and the second acoustic virial coefficients $\beta_{\mathrm{a}}(T)$ determined from the speed of sound data in this work. The exact equation that relates the virial coefficients [36] is:

$$\beta_{\mathrm{a}} = 2B + 2\left(\gamma^{\mathrm{pg}} - 1\right) T \frac{dB}{dT} + \frac{\left(\gamma^{\mathrm{pg}} - 1\right)^2}{\gamma^{\mathrm{pg}}} T^2 \frac{d^2 B}{dT^2} \tag{14}$$

To perform the calculation, the $C_{\mathrm{p,m}}^{\mathrm{pg}}$ data is regressed to the simple form:

$$\frac{C_{\mathrm{p,m}}^{\mathrm{pg}}\left(T\right)}{R} = u_0 + u_1 \frac{\left(v_1 / T\right)^2 e^{v_1/T}}{\left(e^{v_1/T} - 1\right)^2} \tag{15}$$

which is sufficient for the residuals of the fit to fall within experimental uncertainty for the narrow measurement range employed in this research. Two different effective intermolecular potentials are thus used to represent $B(T)$, the hard-core square well potential (HCSW):

$$\begin{aligned} U\left(r\right) &= \infty \quad , \quad r < \sigma_{\mathrm{SW}} \\ U\left(r\right) &= -\varepsilon_{\mathrm{SW}} \quad , \quad \sigma_{\mathrm{SW}} \le r \le g\sigma_{\mathrm{SW}} \\ U\left(r\right) &= 0 \quad , \quad r > g\sigma_{\mathrm{SW}} \end{aligned} \tag{16}$$

where $\sigma_{\mathrm{SW}}$ is the hard-core length, $\varepsilon_{\mathrm{SW}}$ is the well depth, and $g$ is $\sigma_{\mathrm{SW}}$ times the length of the square well; and the Lennard-Jonnes (12,6) potential (LJ (12,6)):

$$U\left(r\right) = 4\varepsilon_{\mathrm{LJ}} \left[ \left( \frac{\sigma_{\mathrm{LJ}}}{r} \right)^{12} - \left( \frac{\sigma_{\mathrm{LJ}}}{r} \right)^{6} \right] \tag{17}$$

where $\varepsilon_{\mathrm{LJ}}$ is the depth of the potential well and $\sigma_{\mathrm{LJ}}$ is the separation at which $U(r) = -\varepsilon_{\mathrm{LJ}}$. According to the pairwise non-polar spherically symmetric relation of $B(T)$ with the potential energy function $U(r)$:

$$B\left(T\right) = -2\pi \int_0^\infty \left[ e^{-U(r)/k_{\mathrm{B}}T} - 1 \right] r^2 dr \tag{18}$$

the HCSW potential yields:

$$B = \left( \frac{2\pi N_{\mathrm{a}} \sigma_{\mathrm{SW}}^3}{3} \right) \left[ g^3 - \left(g^3 - 1\right) e^{\varepsilon_{\mathrm{SW}}/k_{\mathrm{B}}T} \right] = a + b e^{c/T} \tag{19}$$

$$T\frac{dB}{dT} = -\frac{c}{T}be^{c/T} \tag{20}$$

$$T^2\frac{d^2B}{dT^2} = \left[\frac{2c}{T} + \frac{c^2}{T^2}\right]b\,\mathrm{e}^{c/T} \tag{21}$$

and the LJ (12,6) potential, expressed in an analytical closed form in terms of a linear combination of the modified Bessel functions of the first kind $I$ [46], yields:

$$B^*\left(T^*\right) = \frac{\sqrt{2}\pi}{2T^*}e^{1/2T^*}\left(I_{3/4}\left(\frac{1}{2T^*}\right) + I_{-3/4}\left(\frac{1}{2T^*}\right) - I_{1/4}\left(\frac{1}{2T^*}\right) - I_{-1/4}\left(\frac{1}{2T^*}\right)\right) \tag{22}$$

$$T^*\frac{dB^*\left(T^*\right)}{dT^*} = -\frac{\sqrt{2}\pi}{8T^*}e^{1/2T^*}\left(I_{3/4}\left(\frac{1}{2T^*}\right) + I_{-3/4}\left(\frac{1}{2T^*}\right) - 3I_{1/4}\left(\frac{1}{2T^*}\right) - 3I_{-1/4}\left(\frac{1}{2T^*}\right)\right) \tag{23}$$

$$T^{*2}\frac{d^2B^*\left(T^*\right)}{dT^{*2}} = \frac{\pi}{16\sqrt{2}T^{*2}}e^{1/2T^*}\begin{bmatrix}\left(-4+5T^*\right)\left(I_{3/4}\left(\frac{1}{2T^*}\right) + I_{-3/4}\left(\frac{1}{2T^*}\right)\right) \\ -\left(4+21T^*\right)\left(I_{1/4}\left(\frac{1}{2T^*}\right) - I_{-1/4}\left(\frac{1}{2T^*}\right)\right)\end{bmatrix} \tag{24}$$

where $T^* = k_BT/\varepsilon_{LJ}$ and $B^*(T^*) = B(T)/(2\pi N_a\sigma_{LJ}^3/3)$. Combining equation (15) and equations (19) to (21) with expression (14) is used to perform the non-linear regression of the experimental $\beta_a(T)$ data for each mixture, assuming that the hydrogen content is low enough and the temperature high enough to neglect the effect of the quantum corrections of hydrogen to $B(T)$. The procedure is performed with two different effective intermolecular potentials to ensure the robustness of the calculation, given that $B(T)$ is not very sensitive to their shape.

Finally, the methane-hydrogen interaction second virial coefficient $B_{12}(T)$ is determined from $B(T,x)$:

$$B\left(T,x\right) = x_1^2B_{11}\left(T\right) + 2x_1x_2B_{12}\left(T\right) + x_2^2B_{22}\left(T\right) \tag{25}$$

where $B_{11}(T)$ and $B_{22}(T)$ are the pure methane and hydrogen second density virial coefficients from their reference equations of state [17] and [18], respectively, and $x_1$ and $x_2$ are the amount of substance (mole fraction) of methane and hydrogen.

## 4. Measurement uncertainty

Speed of sound uncertainty $u(w)$ is evaluated applying the law of propagation of uncertainties in accordance with the Guide to the Expression of Uncertainty in Measurement (GUM) [47] to equation (2), with the inclusion of the uncertainties of the thermodynamic state (uncertainties of temperature $u(T)$, pressure $u(p)$, and composition of the mixture $u(M)$):

$$u(w)=\left[u_{\mathrm{Eq.2}}(w)^2+\left(\left(\frac{\partial w}{\partial T}\right)_{p,M}u(T)\right)^2+\left(\left(\frac{\partial w}{\partial p}\right)_{T,M}u(p)\right)^2+\left(\left(\frac{\partial w}{\partial M}\right)_{T,p}u(M)\right)^2\right]^{1/2} \quad (26)$$

$$u_{\mathrm{Eq.2}}(w)=\left[\left(\left(\frac{2\pi}{n}\sum_{\mathrm{i=1}}^{\mathrm{n}}\frac{f_{\mathrm{i}}}{\nu_{\mathrm{i}}}\right)u(a)\right)^2+\left(\left(\frac{2\pi a}{n}\sum_{\mathrm{i=1}}^{\mathrm{n}}\frac{1}{\nu_{\mathrm{i}}}\right)u(f)\right)^2+\left(u(w)_{\mathrm{disp}}\right)^2\right]^{1/2} \quad (27)$$

where $u(a)$ is the uncertainty of the internal radius, $u(f)$ is the uncertainty of the fitting of the resonance frequency to equation (1), and $u(w)_{\mathrm{disp}}$ is the uncertainty due to the dispersion of the $n$ acoustic modes considered when determining $w(p,T)$ in line with the above discussion. Table 4 displays the uncertainty contributions considered in the uncertainty of speed of sound for the binary ($CH_4$ + $H_2$) mixtures in this work. The expanded ($k$ = 2) relative uncertainty in speed of sound $U_{\mathrm{r}}(w)$ is 220 parts in $10^6$, with the inner radius uncertainty being the most significant term.

Uncertainties of the derived properties are estimated by the GUM procedure [47] from equations (9) to (11) and the uncertainty of the regression coefficients of the polynomial equation to the square speed of sound reported in table 3:

$$u(\gamma^{\mathrm{pg}})=\left[\left(\left(\frac{A_0}{RT}\right)u(M)\right)^2+\left(\left(\frac{MA_0}{R^2T}\right)u(R)\right)^2+\left(\left(\frac{MA_0}{RT^2}\right)u(T)\right)^2+\left(\left(\frac{M}{RT}\right)u(A_0)\right)^2\right]^{1/2} \quad (28)$$

$$u(C_{\mathrm{V,m}}^{\mathrm{pg}})=\left[\left(\left(\frac{1}{\gamma^{\mathrm{pg}}-1}\right)u(R)\right)^2+\left(\left(\frac{R}{\left[\gamma^{\mathrm{pg}}-1\right]^2}\right)u(\gamma^{\mathrm{pg}})\right)^2\right]^{1/2} \quad (29)$$

$$u(C_{\mathrm{p,m}}^{\mathrm{pg}})=\left[\left(C_{\mathrm{V,m}}^{\mathrm{pg}}u(\gamma^{\mathrm{pg}})\right)^2+\left(\gamma^{\mathrm{pg}}u(C_{\mathrm{V,m}}^{\mathrm{pg}})\right)^2\right]^{1/2} \quad (30)$$

$$u(\beta_a) = \left[ \left( \frac{RT}{A_0} u(A_1) \right)^2 + \left( \frac{A_1 RT}{A_0^2} u(A_0) \right)^2 + \left( \frac{A_1 T}{A_0} u(R) \right)^2 + \left( \frac{A_1 R}{A_0} u(T) \right)^2 \right]^{1/2} \quad (31)$$

$$u(\gamma_a) = \left[ \begin{array}{c} \left(\beta_a u(B)\right)^2 + \left(Bu(\beta_a)\right)^2 + \left( \frac{(RT)^2}{A_0} u(A_2) \right)^2 \\ + \left( \frac{A_2 (RT)^2}{A_0^2} u(A_0) \right)^2 + \left( \frac{2A_2 RT^2}{A_0} u(R) \right)^2 + \left( \frac{2A_2 R^2 T}{A_0} u(T) \right)^2 \end{array} \right]^{1/2} \quad (32)$$

where $u(R)$ is obtained from the recommended value of [48]. The expanded ($k$ = 2) relative uncertainty of $\gamma^{pg}$ is better than 0.02 %, below 0.1 % for $C_{v,m}^{pg}$ and $C_{p,m}^{pg}$, and between (1 to 8) % and (2 to 9) % for $\beta_a$ and $\gamma_a$, respectively.

**Table 4.** Uncertainty budget for the speed of sound *w* measurements. Unless otherwise specified, uncertainty *u* is indicated with a coverage factor *k* = 1.

| Source | Magnitude | | Contribution to speed of sound uncertainty, $10^6 \cdot u_r(w)$ / (m·s$^{-1}$)/(m·s$^{-1}$) |
|---|---|---|---|
| Temperature | Calibration | 0.0020 K | |
| | Resolution | $7.2\cdot10^{-7}$ K | |
| | Repeatability | $4.2\cdot10^{-5}$ K | |
| | Gradient (across hemispheres) | $1.0\cdot10^{-3}$ K | |
| | Sum | 0.0023 K | 3.4 |
| Pressure | Calibration | $(7.5\cdot10^{-5}\cdot p + 2\cdot10^{-4})$ MPa | |
| | Resolution | $2.9\cdot10^{-5}$ MPa | |
| | Repeatability | $2.3\cdot10^{-5}$ MPa | |
| | Sum | (1.2 to 8.3)$\cdot10^{-4}$ MPa | 4.8 |
| Gas composition | Purity | $7.5\cdot10^{-7}$ kg·mol$^{-1}$ | |
| | Molar mass | $1.7\cdot10^{-7}$ kg·mol$^{-1}$ | |
| | Sum | $7.7\cdot10^{-7}$ kg·mol$^{-1}$ | 27 |
| Radius from speed of sound in Ar | Temperature | $1.5\cdot10^{-9}$ m | |
| | Pressure | $1.6\cdot10^{-10}$ m | |
| | Gas Composition | $4.1\cdot10^{-9}$ m | |

| | | |
|---|---|---|
| Frequency fitting | $4.9·10^{-7}$ m | |
| Regression | $1.7·10^{-6}$ m | |
| Equation of State | $2.3·10^{-6}$ m | |
| Dispersion of modes | $2.9·10^{-6}$ m | |
| Sum | $4.2·10^{-6}$ m | 99 |
| Frequency fitting | 0.024 Hz | 1.4 |
| Dispersion of modes | $1.8·10^{-2}$ m·s$^{-1}$ | 33 |
| Sum of all contributions to $w$ | | 110 |
| $10^6·U_r(w)$ / (m·s$^{-1}$)/(m·s$^{-1}$) (*) | | 220 |

(*) Uncertainty with coverage factor $k$ = 2.

## 5. Results

Experimental speed of sound $w(p, T)$ data from the corrected resonance frequencies for the three binary ($CH_4 + H_2$) mixtures measured in this work are listed in tables 5, 6, and 7, together with the speed of sound determined from reference AGA8-DC92 [10] and GERG-2008 EoS [8] and the relative deviations of the experimental data of this work from the reference equations of state. Expanded ($k$ = 2) relative uncertainty in speed of sound $U_r(w)$ is reported in detail in table 4, as described above. Data comprise results at temperatures $T$ = (273.16, 300, 325, 350, and 375) K in the pressure range $p$ = (0.5 to 20) MPa.

**Table 5.** Experimental speed of sound $w_{exp}$ with their corresponding relative expanded ($k$ = 2) uncertainties(*) and comparison with EoS GERG-2008 [8] and AGA8-DC92 [10] for the (0.95 $CH_4$ + 0.05 $H_2$) mixture with the composition specified in Table 1.

| $p$ / MPa | $w_{exp}$ / m·s$^{-1}$ | $10^6·\Delta w_{AGA}$(**) | $10^6·\Delta w_{GERG}$(**) | $p$ / MPa | $w_{exp}$ / m·s$^{-1}$ | $10^6·\Delta w_{AGA}$(**) | $10^6·\Delta w_{GERG}$(**) |
|---|---|---|---|---|---|---|---|
| $T$ = 273.16 K | | | | $T$ = 300.00 K | | | |
| 0.48093 | 439.555 | −57 | −111 | 0.49428 | 459.595 | −102 | −130 |
| 0.95352 | 437.751 | 11 | −61 | 0.98708 | 458.435 | −13 | −63 |
| 1.94476 | 434.234 | 113 | 2 | 1.97706 | 456.333 | 55 | −34 |
| 2.95350 | 431.122 | 209 | 68 | 2.98075 | 454.633 | 134 | 18 |
| 3.96005 | 428.587 | 306 | 140 | 3.98305 | 453.416 | 214 | 76 |
| 4.96989 | 426.732 | 406 | 217 | 4.98398 | 452.741 | 300 | 141 |
| 5.97188 | 425.699 | 512 | 291 | 5.97553 | 452.661 | 389 | 203 |
| 6.99156 | 425.612 | 637 | 353 | 6.98800 | 453.252 | 499 | 272 |
| 7.99623 | 426.620 | 779 | 399 | 7.98929 | 454.552 | 601 | 319 |
| 8.98598 | 428.798 | 924 | 410 | 8.99078 | 456.625 | 710 | 357 |
| 9.98461 | 432.303 | 1070 | 394 | 9.97066 | 459.438 | 810 | 374 |
| 10.98657 | 437.223 | 1193 | 349 | 10.79807 | 462.444 | 887 | 378 |
| 11.80487 | 442.323 | 1272 | 308 | 11.48854 | 465.407 | 948 | 381 |
| 12.49109 | 447.348 | 1307 | 263 | 12.29517 | 469.388 | 994 | 368 |
| 13.27182 | 453.873 | 1297 | 191 | 12.97841 | 473.204 | 1022 | 359 |
| 14.00158 | 460.724 | 1256 | 116 | 13.81568 | 478.424 | 1032 | 345 |
| 14.79487 | 468.949 | 1193 | 42 | 14.51408 | 483.224 | 1013 | 327 |
| 15.50366 | 476.905 | 1085 | −63 | 15.30042 | 489.101 | 976 | 313 |
| 16.31147 | 486.642 | 1016 | −124 | 15.98613 | 494.622 | 943 | 317 |
| 17.04491 | 495.995 | 940 | −186 | 16.55422 | 499.447 | 888 | 303 |
| 17.83195 | 506.482 | 834 | −274 | 17.49614 | 507.934 | 786 | 285 |
| 18.51570 | 515.940 | 784 | −300 | 18.29560 | 515.566 | 677 | 255 |
| 19.32866 | 527.447 | 611 | −428 | 18.93451 | 521.922 | 585 | 224 |
| 19.99010 | 536.983 | 414 | −572 | | | | |
| | | $T$ = 325.00 K | | | | $T$ = 350.00 K | |
| 0.48110 | 476.861 | −319 | −348 | 0.47315 | 493.060 | −399 | −444 |
| 0.93346 | 476.238 | −231 | −285 | 1.46058 | 492.495 | −242 | −358 |
| 1.97447 | 475.041 | −95 | −200 | 3.17997 | 492.204 | −85 | −285 |
| 2.97418 | 474.257 | 0 | −139 | 5.43052 | 493.408 | 136 | −131 |
| 3.97804 | 473.877 | 93 | −72 | 7.44669 | 496.149 | 333 | 7 |
| 4.97810 | 473.935 | 181 | −8 | 9.48960 | 500.664 | 548 | 144 |
| 5.98867 | 474.472 | 272 | 58 | 11.49720 | 506.859 | 717 | 224 |
| 6.98692 | 475.502 | 360 | 112 | 13.52754 | 514.987 | 950 | 390 |
| 8.00016 | 477.092 | 457 | 169 | 15.52198 | 524.663 | 1036 | 463 |
| 8.98548 | 479.182 | 547 | 210 | 17.53712 | 536.081 | 1033 | 528 |
| 9.99482 | 481.917 | 667 | 272 | 19.29931 | 547.308 | 966 | 593 |
| 10.99361 | 485.201 | 749 | 295 | | | | |
| 11.69968 | 487.887 | 810 | 316 | | *T* = 375.00 K | | |
| 12.50938 | 491.329 | 858 | 323 | 0.48389 | 508.279 | −592 | −662 |
| 13.29378 | 495.039 | 896 | 332 | 1.46596 | 508.242 | −507 | −660 |
| 13.98088 | 498.584 | 911 | 331 | 2.42463 | 508.491 | −372 | −591 |
| 14.76786 | 502.989 | 931 | 348 | 4.44704 | 509.882 | −133 | −447 |
| 15.49379 | 507.357 | 923 | 353 | 6.43509 | 512.456 | 51 | −336 |
| 16.26660 | 512.321 | 895 | 354 | 8.47427 | 516.501 | 351 | −96 |
| 16.99333 | 517.282 | 869 | 370 | 10.46544 | 521.750 | 537 | 17 |
| 17.77873 | 522.940 | 825 | 388 | 12.51196 | 528.583 | 744 | 169 |
| 18.48426 | 528.281 | 794 | 423 | 14.49418 | 536.523 | 864 | 255 |
| 19.26238 | 534.431 | 755 | 468 | 16.49553 | 545.858 | 968 | 379 |
| 19.81381 | 538.946 | 732 | 506 | 18.32910 | 555.480 | 1009 | 484 |

(*) Expanded uncertainties ($k = 2$): $U(p) = 7.5\cdot10^{-5}$ ($p$/Pa) + 200 Pa; $U(T)$ = 4 mK; $U_r(w) = 2.2\cdot10^{-4}$ $m\cdot s^{-1}$/ $m\cdot s^{-1}$.

(**) $\Delta w_{AGA} = (w_{exp} - w_{AGA})/w_{AGA}$; $\Delta w_{GERG} = (w_{exp} - w_{GERG})/w_{GERG}$

**Table 6.** Experimental speed of sound $w_{exp}$ with their corresponding relative expanded ($k$ = 2) uncertainties$^{(*)}$ after applying the acoustic model and data reduction, and comparison with EoS GERG-2008 [8] and AGA8-DC92 [10] for the (0.90 $CH_4$ + 0.10 $H_2$) mixture with the composition specified in Table 1.

| $p$ / MPa | $w_{exp}$ / $m \cdot s^{-1}$ | $10^6 \cdot \Delta w_{AGA}$$^{(**)}$ | $10^6 \cdot \Delta w_{GERG}$$^{(**)}$ | $p$ / MPa | $w_{exp}$ / $m \cdot s^{-1}$ | $10^6 \cdot \Delta w_{AGA}$$^{(**)}$ | $10^6 \cdot \Delta w_{GERG}$$^{(**)}$ |
|---|---|---|---|---|---|---|---|
| | $T$ = 273.16 K | | | | $T$ = 300.00 K | | |
| 0.48194 | 450.857 | −184 | −272 | 0.47257 | 471.450 | −168 | −218 |
| 0.97037 | 449.306 | −76 | −206 | 0.96683 | 470.549 | −51 | −126 |
| 1.96211 | 446.443 | 52 | −93 | 1.95056 | 468.971 | 47 | −60 |
| 2.93736 | 444.065 | 169 | −2 | 2.96143 | 467.777 | 165 | 40 |
| 3.97724 | 442.109 | 269 | 98 | 3.96670 | 467.050 | 266 | 131 |
| 4.97151 | 440.89 | 386 | 190 | 4.97623 | 466.832 | 362 | 214 |
| 5.98632 | 440.405 | 511 | 273 | 5.98327 | 467.184 | 478 | 301 |
| 6.99823 | 440.782 | 640 | 321 | 6.98692 | 468.134 | 588 | 360 |
| 7.99875 | 442.098 | 788 | 327 | 7.99063 | 469.732 | 707 | 403 |
| 9.00367 | 444.454 | 955 | 285 | 9.00822 | 472.045 | 825 | 412 |
| 10.02703 | 447.999 | 1137 | 200 | 10.00703 | 475.030 | 944 | 402 |
| 10.98665 | 452.421 | 1286 | 82 | 10.99218 | 478.679 | 1039 | 358 |
| 11.78485 | 456.927 | 1374 | −36 | 11.78796 | 482.165 | 1136 | 341 |
| 12.50960 | 461.669 | 1451 | −158 | 12.50687 | 485.706 | 1189 | 298 |
| 13.32894 | 467.762 | 1482 | −305 | 13.27081 | 489.808 | 1071 | 91 |
| 14.04170 | 473.672 | 1463 | −442 | 14.00767 | 494.223 | 1061 | 15 |
| 14.83328 | 480.873 | 1422 | −584 | 14.78674 | 499.314 | 1047 | −48 |
| 15.54584 | 487.885 | 1344 | −714 | 15.46754 | 504.098 | 1007 | −104 |
| 16.33184 | 496.164 | 1259 | −836 | 16.29921 | 510.361 | 955 | −145 |
| 17.02278 | 503.851 | 1135 | −958 | 17.01472 | 516.086 | 884 | −181 |
| 17.84943 | 513.514 | 982 | −1074 | 17.83146 | 522.980 | 785 | −216 |
| 18.52874 | 521.777 | 857 | −1163 | 18.52538 | 529.139 | 728 | −199 |
| 19.36566 | 532.266 | 650 | −1322 | 19.29316 | 536.226 | 649 | −181 |
| 20.19624 | 542.942 | 335 | −1540 | 19.62633 | 539.385 | 614 | −172 |
| | | *T* = 325.00 K | | | | *T* = 350.00 K | |
| 0.48312 | 488.984 | −615 | −659 | 0.44374 | 505.544 | −836 | −889 |
| 0.98432 | 488.534 | −483 | −559 | 1.93669 | 505.381 | −546 | −707 |
| 1.93362 | 487.853 | −365 | −485 | 3.93160 | 506.182 | −332 | −563 |
| 2.97481 | 487.481 | −240 | −387 | 5.94493 | 508.383 | −92 | −369 |
| 3.97419 | 487.523 | −113 | −278 | 7.98214 | 512.128 | 168 | −176 |
| 4.94673 | 487.942 | −25 | −207 | 9.97305 | 517.281 | 331 | −111 |
| 5.95494 | 488.821 | 82 | −122 | 12.01615 | 524.223 | 550 | −6 |
| 6.97224 | 490.172 | 173 | −67 | 13.91129 | 532.096 | 673 | 30 |
| 7.99075 | 492.034 | 291 | −2 | 16.00345 | 542.389 | 797 | 127 |
| 8.97511 | 494.316 | 385 | 28 | 18.01956 | 553.756 | 806 | 215 |
| 9.99517 | 497.209 | 493 | 54 | 19.35150 | 561.986 | 760 | 284 |
| 10.98492 | 500.529 | 584 | 59 | | | | |
| 11.78936 | 503.613 | 666 | 70 | | | *T* = 375.00 K | |
| 12.50291 | 506.620 | 707 | 52 | 0.49781 | 521.072 | −1204 | −1276 |
| 13.32487 | 510.424 | 769 | 52 | 1.93382 | 521.677 | −887 | −1096 |
| 14.00386 | 513.823 | 799 | 44 | 3.94102 | 523.391 | −609 | −928 |
| 14.79585 | 518.074 | 807 | 22 | 5.94221 | 526.260 | −342 | −711 |
| 15.50781 | 522.168 | 822 | 24 | 7.96357 | 530.320 | −170 | −594 |
| 16.29524 | 526.981 | 828 | 40 | 9.94891 | 535.576 | 86 | −401 |
| 17.02539 | 531.692 | 809 | 48 | 11.96639 | 542.120 | 258 | −304 |
| 17.82071 | 537.109 | 811 | 101 | 13.98667 | 549.947 | 468 | −155 |
| 18.51169 | 542.020 | 779 | 135 | 16.00496 | 558.921 | 539 | −79 |
| 19.33709 | 548.121 | 710 | 160 | 18.00068 | 568.943 | 638 | 83 |
| 20.11588 | 554.076 | 574 | 132 | 19.46940 | 576.959 | 640 | 191 |

(*) Expanded uncertainties ($k$ = 2): $U(p) = 7.5\cdot10^{-5}$ ($p$/Pa) + 200 Pa; $U(T)$ = 4 mK; $U_r(w) = 2.2\cdot10^{-4}$ m·s$^{-1}$/ m·s$^{-1}$.

(**) $\Delta w_{AGA} = (w_{exp} - w_{AGA})/w_{AGA}$; $\Delta w_{GERG} = (w_{exp} - w_{GERG})/w_{GERG}$

**Table 7.** Experimental speed of sound $w_{exp}$ with their corresponding relative expanded ($k$ = 2) uncertainties(*) after applying the acoustic model and data reduction, and comparison with EoS GERG-2008 [8] and AGA8-DC92 [10] for the (0.50 $CH_4$ + 0.50 $H_2$) mixture with the composition specified in Table 1.

| $p$ / MPa | $w_{exp}$ / m·s$^{-1}$ | $10^6\cdot\Delta w_{AGA}$(**) | $10^6\cdot\Delta w_{GERG}$(**) | $p$ / MPa | $w_{exp}$ / m·s$^{-1}$ | $10^6\cdot\Delta w_{AGA}$(**) | $10^6\cdot\Delta w_{GERG}$(**) |
|---|---|---|---|---|---|---|---|
| | | $T$ = 273.16 K | | | | $T$ = 300.00 K | |
| 0.45721 | 583.614 | −1066 | −1548 | 0.43980 | 609.770 | −1582 | −1810 |
| 0.95085 | 584.503 | −638 | −1221 | 1.09278 | 611.355 | −1013 | −1366 |
| 1.93270 | 586.262 | −223 | −964 | 3.03834 | 616.062 | −488 | −1098 |
| 2.90623 | 588.220 | −2 | −870 | 5.12235 | 622.049 | −259 | −1088 |
| 3.94951 | 590.645 | 173 | −829 | 7.13158 | 628.994 | 68 | −1010 |
| 4.91880 | 593.214 | 295 | −845 | 9.17651 | 637.026 | 203 | −1179 |
| 5.95601 | 596.418 | 582 | −737 | 11.11824 | 645.659 | 366 | −1318 |
| 6.95309 | 599.744 | 679 | −845 | 13.15263 | 655.647 | 468 | −1486 |
| 7.97987 | 603.595 | 889 | −879 | 15.15777 | 666.391 | 501 | −1599 |
| 8.97883 | 607.646 | 1021 | −1011 | 17.17939 | 678.040 | 413 | −1643 |
| 9.98797 | 612.083 | 1156 | −1160 | 19.28692 | 690.989 | 185 | −1561 |
| 10.99779 | 616.861 | 1287 | −1320 | | | | |
| 11.75580 | 620.707 | 1449 | −1372 | | | $T$ = 325.00 K | |
| 12.45490 | 624.386 | 1543 | −1463 | 0.47781 | 632.470 | −2446 | −2597 |
| 13.26532 | 628.839 | 1642 | −1562 | 1.47888 | 635.213 | −1810 | −2122 |
| 13.94759 | 632.736 | 1707 | −1641 | 3.45866 | 640.871 | −1253 | −1767 |
| 14.65163 | 636.907 | 1774 | −1699 | 5.45123 | 647.180 | −1114 | −1770 |
| 15.50359 | 642.123 | 1807 | −1775 | 7.44679 | 654.364 | −966 | −1758 |
| 16.30589 | 647.214 | 1816 | −1824 | 9.47230 | 662.418 | −947 | −1879 |
| 17.00825 | 651.820 | 1821 | −1827 | 11.45379 | 671.099 | −913 | −1963 |
| 17.81280 | 657.284 | 1861 | −1748 | 13.46500 | 680.807 | −683 | −1796 |
| 18.51126 | 662.098 | 1790 | −1739 | 15.47644 | 691.054 | −742 | −1811 |
| 19.12728 | 666.438 | 1714 | −1708 | 17.51623 | 702.078 | −909 | −1777 |
| 20.01618 | 672.844 | 1576 | −1634 | 19.41718 | 712.964 | −1054 | −1555 |

(*) Expanded uncertainties ($k$ = 2): $U(p) = 7.5 \cdot 10^{-5}$ ($p$/Pa) + 200 Pa; $U(T)$ = 4 mK; $U_r(w) = 2.2 \cdot 10^{-4}$ m·s$^{-1}$/ m·s$^{-1}$.

(**) $\Delta w_{AGA} = (w_{exp} - w_{AGA})/w_{AGA}$; $\Delta w_{GERG} = (w_{exp} - w_{GERG})/w_{GERG}$

As regards the derived properties, adiabatic coefficient $\gamma^{pg}$, molar isochoric heat capacity $C_{v,m}{}^{pg}$, molar isobaric heat capacity $C_{p,m}{}^{pg}$, acoustic second virial coefficient $\beta_a$ and acoustic third virial coefficient $\gamma_a$, results are reported in table 8 with their corresponding expanded ($k$ = 2) uncertainties and are compared to AGA8-DC92 [10] and GERG-2008 EoS [8]. In addition, the density second virial coefficients obtained for the two ($CH_4$ + $H_2$) mixtures with lower hydrogen content in this work are depicted in figures 8 and 9, respectively, and are discussed below, and the results are compared with the literature and established reference EoS values.

**Table 8.** Adiabatic coefficient $\gamma^{pg}$, isobaric heat capacity $C_{p,m}{}^{pg}$, acoustic second virial coefficient $\beta_a$, and acoustic third virial coefficient $\gamma_a$ derived from the speed of sound data with their corresponding relative expanded ($k$ = 2) uncertainty $U_r$, and comparison with AGA8-DC92 [10] and GERG-2008 EoS [8]. The superscript *pg* indicates perfect-gas property.

| $T$ / K | $\gamma^{pg}$ | $10^2 \cdot U_r(\gamma^{pg})$ | $10^2 \cdot \Delta\gamma^{pg}{}_{AGA}$ | $10^2 \cdot \Delta\gamma^{pg}{}_{GERG}$ | $C_{p,m}{}^{pg}$ / $J \cdot mol^{-1} \cdot K^{-1}$ | $10^2 \cdot U_r(C_{p,m}{}^{pg})$ | $10^2 \cdot \Delta C_p{}^{pg}{}_{AGA}$ | $10^2 \cdot \Delta C_p{}^{pg}{}_{GERG}$ |
|---|---|---|---|---|---|---|---|---|
| | | | | ($0.95\ CH_4 + 0.05\ H_2$) | | | | |
| 273.16 | 1.31684 | 0.018 | 0.0087 | 0.0021 | 34.556 | 0.080 | −0.027 | −0.0073 |
| 300.00 | 1.30628 | 0.017 | −0.027 | −0.027 | 35.462 | 0.073 | 0.088 | 0.088 |
| 325.00 | 1.29493 | 0.015 | −0.081 | −0.080 | 36.505 | 0.071 | 0.28 | 0.27 |
| 350.00 | 1.28340 | 0.020 | −0.10 | −0.10 | 37.653 | 0.096 | 0.36 | 0.36 |
| 375.00 | 1.27189 | 0.019 | −0.11 | −0.11 | 38.894 | 0.093 | 0.40 | 0.41 |
| | | | | ($0.90\ CH_4 + 0.10\ H_2$) | | | | |
| 273.16 | 1.31972 | 0.017 | −0.056 | −0.069 | 34.320 | 0.070 | 0.17 | 0.21 |
| 300.00 | 1.30992 | 0.015 | −0.048 | −0.051 | 35.142 | 0.068 | 0.16 | 0.16 |
| 325.00 | 1.29811 | 0.015 | −0.16 | −0.16 | 36.206 | 0.066 | 0.52 | 0.52 |
| 350.00 | 1.28666 | 0.020 | −0.19 | −0.19 | 37.319 | 0.096 | 0.66 | 0.66 |
| 375.00 | 1.27495 | 0.019 | −0.23 | −0.23 | 38.555 | 0.093 | 0.83 | 0.84 |
| | | | | ($0.50\ CH_4 + 0.50\ H_2$) | | | | |
| 273.16 | 1.35072 | 0.013 | −0.24 | −0.31 | 32.021 | 0.053 | 0.68 | 0.88 |
| 300.00 | 1.34159 | 0.018 | −0.34 | −0.36 | 32.655 | 0.073 | 1.0 | 1.1 |
| 325.00 | 1.33093 | 0.020 | −0.54 | −0.55 | 33.439 | 0.078 | 1.6 | 1.7 |
| | $10^7 \cdot \beta_a$ / $m^3 \cdot mol^{-1}$ | $10^2 \cdot U_r(\beta_a)$ | $10^2 \cdot \Delta\beta_{a,AGA}$ | $10^2 \cdot \Delta\beta_{a,\ GERG}$ | $10^{10} \cdot \gamma_a$ / $(m^3 \cdot mol^{-1})^2$ | $10^2 \cdot U_r(\gamma_a)$ | $10^2 \cdot \Delta\gamma_{a,AGA}$ | $10^2 \cdot \Delta\gamma_{a,GERG}$ |
| | | | | ($0.95\ CH_4 + 0.05\ H_2$) | | | | |
| 273.16 | −434.1 | 0.013 | 2.4 | 2.7 | 82.0 | 7.6 | 34 | - |
| 300.00 | −286.7 | 1.2 | −1.6 | −0.8 | 57.4 | 4.9 | 0.070 | - |
| 325.00 | −181.5 | 1.8 | −5.8 | −3.8 | 49.9 | 5.6 | −9.3 | - |
| 350.00 | −97.4 | 4.3 | −12.5 | −7.8 | 45.0 | 6.2 | −16 | - |
| 375.00 | −43.1 | 6.0 | −0.1 | 21.4 | 56.3 | 1.8 | 8.3 | - |
| | | | | ($0.90\ CH_4 + 0.10\ H_2$) | | | | |
| 273.16 | −344.1 | 1.0 | −2.6 | −1.8 | 55.1 | 5.1 | −2.8 | - |
| 300.00 | −221.2 | 1.4 | −4.2 | −2.7 | 49.2 | 4.9 | −8.0 | - |
| 325.00 | −121.7 | 1.6 | −13 | −9.7 | 40.8 | 2.7 | −21 | - |
| 350.00 | −50.0 | 8.0 | −23 | −14 | 43.2 | 6.0 | −13 | - |
| 375.00 | 2.3 | 102 | −231 | −65 | 51.8 | 1.7 | 6 | - |
| | | | | ($0.50\ CH_4 + 0.50\ H_2$) | | | | |
| 273.16 | 107.5 | 1.2 | 30 | 13 | 22.1 | 2.8 | −23 | - |
| 300.00 | 166.5 | 1.9 | 21 | 11 | 19.2 | 9.2 | −31 | - |
| 325.00 | 218.6 | 1.6 | 21 | 13 | 12.7 | 17 | −54 | - |

(*) $\Delta X_{\mathrm{EoS}} = (X_{\mathrm{exp}} - X_{\mathrm{EoS}})/X_{\mathrm{EoS}}$ with $X = \gamma^{\mathrm{pg}}$, $C_p^{\mathrm{pg}}$, $\beta_{\mathrm{a}}$, and $\gamma_{\mathrm{a}}$, and EoS = AGA8-DC92 [10], GERG-2008 [8].

## 6. Discussion

Figures 5, 6, and 7 show the relative deviations of $w(p,T)$ from the AGA8-DC92 [10] and GERG-2008 models [8] for binary ($0.95\ CH_4 + 0.05\ H_2$), ($0.90\ CH_4 + 0.10\ H_2$), and ($0.50\ CH_4 + 0.50\ H_2$) mixtures, respectively. Differences between experimental data and the two models are within the experimental uncertainty of $U_{\mathrm{r}}(w) = 220$ parts in $10^6$ (0.022 %) at $T$ = (273.16, 300, and 325) K and pressures below 8 MPa for the mixture with $x_{H_2}$ = 0.05 and at $T$ = (273.16, and 300) K and pressures below 6 MPa for the mixture with $x_{H_2}$ = 0.10. Moreover, deviations agree with the speed of sound uncertainty of the models of 0.2 % for AGA8-DC92 EoS [10] and 0.1 % for GERG-2008 EoS [8] for all the measured points, except for the comparison with the GERG-2008 EoS [8] for the mixture with $x_{H_2}$ = 0.50. Whereas for the ($0.50\ CH_4 + 0.50\ H_2$) mixture, differences compared to the AGA8-DC92 model [10] change from positive values to negative ones as the pressure is reduced and remain within the 0.2 % uncertainty of the EoS, discrepancies from the GERG-2008 EoS [8] always follow a negative sinusoidal-shaped curve outside the 0.1% model uncertainty. The GERG-2008 EoS [8] was expected to perform better than the AGA8-DC92 EoS [10] for this mixture given that its composition lies within the composition range of $x_{H_2}$ = (0.15 to 0.75) of the data used to fit the binary specific departure function of methane + hydrogen. Moreover, the temperature dependence of the discrepancies becomes more pronounced for higher concentrations of hydrogen.

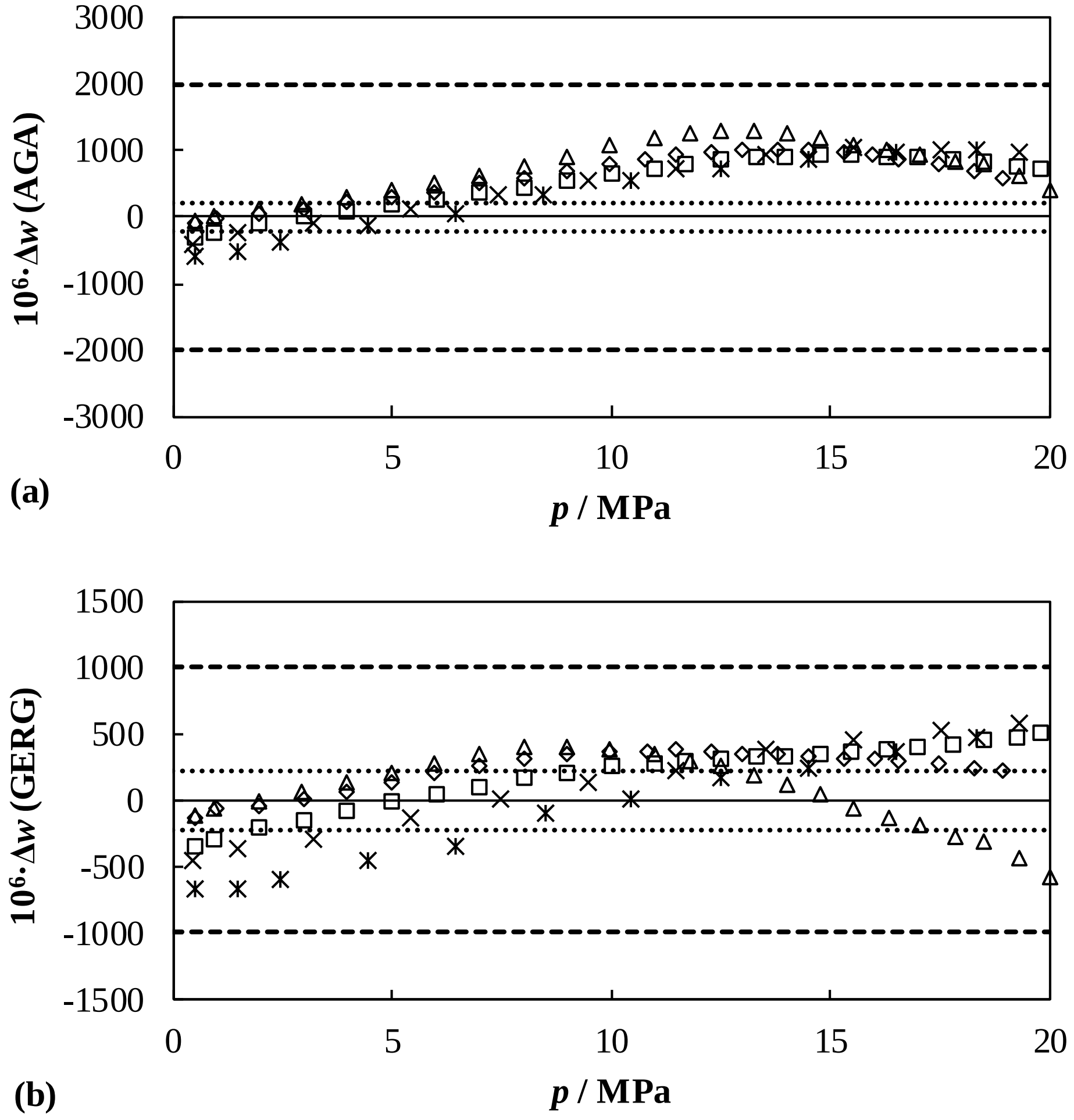

**Figure 5.** Relative deviations $\Delta w = (w_{\mathrm{exp}} - w_{\mathrm{EoS}})/w_{\mathrm{EoS}}$ as function of pressure for binary mixture (0.95 $CH_4$ + 0.05 $H_2$) from calculated values from: (a): AGA8-DC92 EoS [10] and (b): GERG-2008 EoS [8], at temperatures: △ 273.16 K, ◇ 300 K, □ 325 K, × 350 K, ✳ 375 K. Dotted line

represents the expanded ($k$ = 2) experimental uncertainty in speed of sound and dashed line the uncertainty of the EoS.

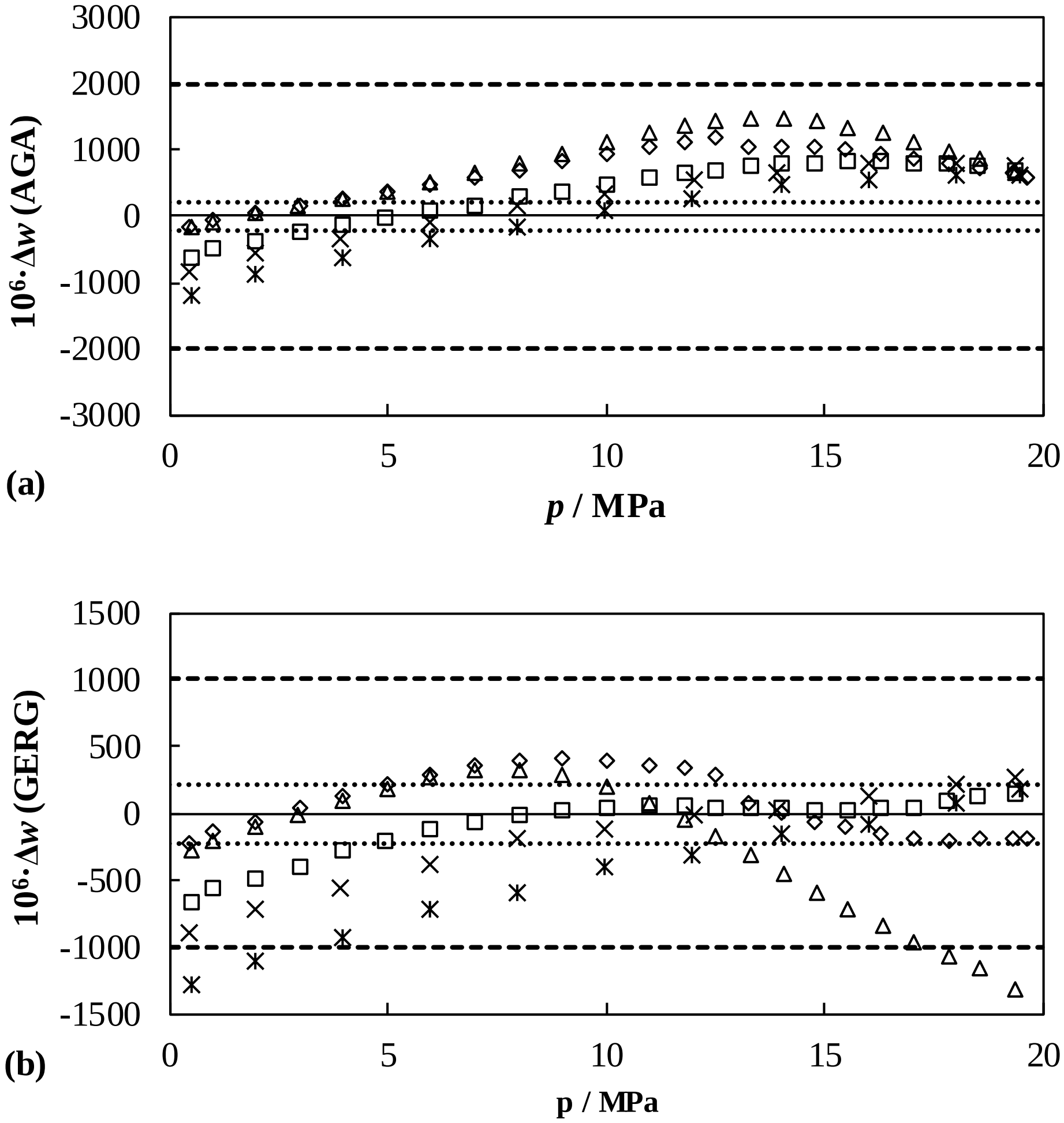

**Figure 6.** Relative deviations $\Delta w = (w_{exp} - w_{EoS})/w_{EoS}$ as function of pressure for binary mixture (0.90 $CH_4$ + 0.10 $H_2$) from calculated values from: (a): AGA8-DC92 EoS [10] and (b): GERG-2008 EoS [8], at temperatures: △ 273.16 K, ◇ 300 K, □ 325 K, × 350 K, ✳ 375 K. Dotted line represents the expanded ($k$ = 2) experimental uncertainty in speed of sound and dashed line the uncertainty of the EoS.

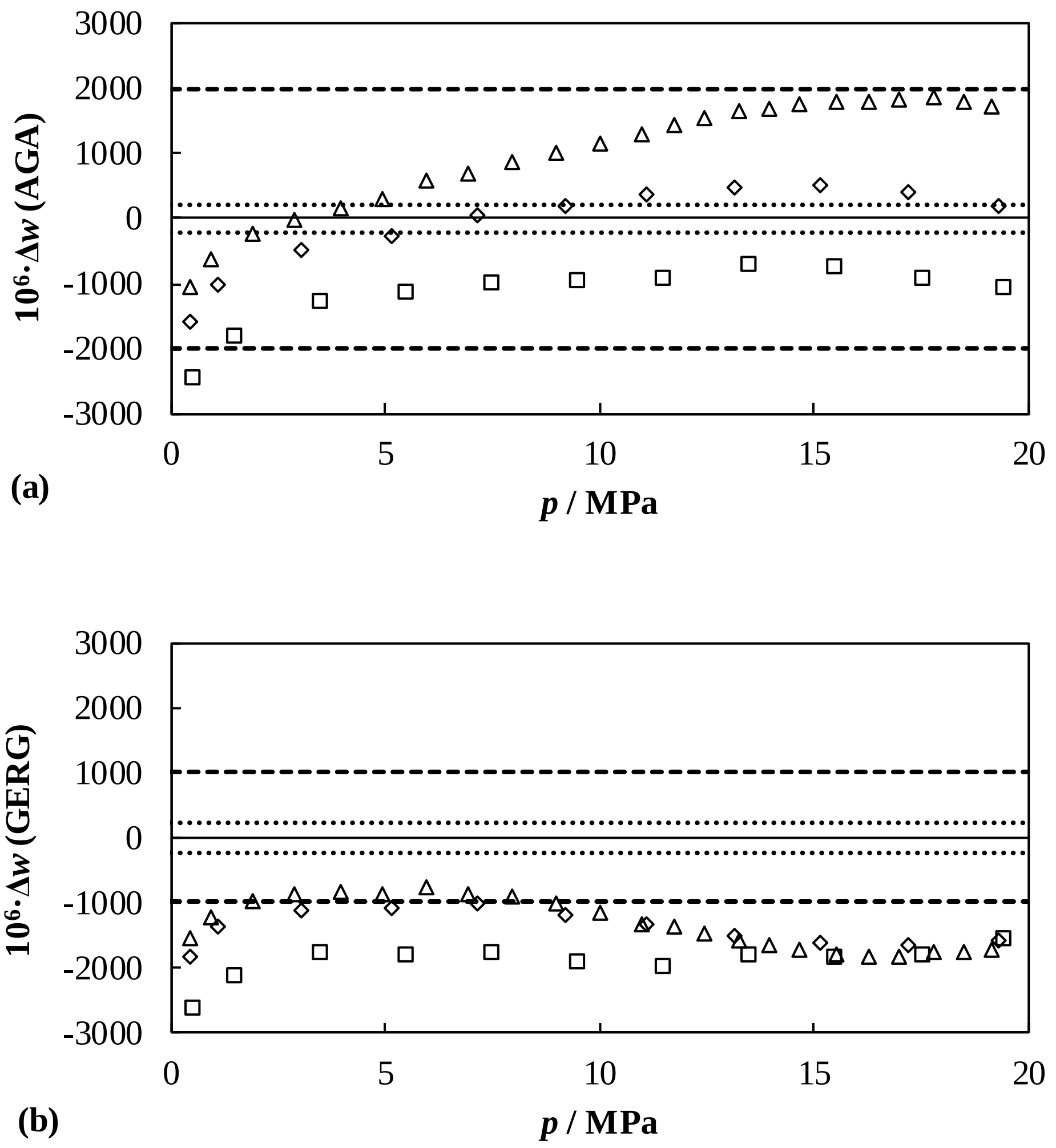

**Figure 7.** Relative deviations $\Delta w = (w_{\mathrm{exp}} - w_{\mathrm{EoS}})/w_{\mathrm{EoS}}$ as function of pressure for binary mixture (0.50 $CH_4$ + 0.50 $H_2$) from calculated values from: (a): AGA8-DC92 EoS [10] and (b): GERG-2008 EoS [8], at temperatures: △ 273.16 K, ◇ 300 K, □ 325 K. Dotted line represents the expanded ($k$ = 2) experimental uncertainty in speed of sound and dashed line the uncertainty of the EoS.

Trends for the (0.95 $CH_4$ + 0.05 $H_2$) mixture are similar to the (0.90 $CH_4$ + 0.10 $H_2$) mixture for both EoS. GERG-2008 [8] yields better results than AGA8-DC92 [10], with an average absolute relative deviation (AAD) of 0.03 % and which are half those of AGA8-DC92 EoS [10] (0.06%). However, AGA8-DC92 EoS [10] is more satisfactory than GERG-2008 EoS [8] for the (0.50 $CH_4$ + 0.50 $H_2$) mixture, with AAD of 0.095% and 0.150%, respectively. Although the trends are similar, the higher the hydrogen content in the mixture, the worse the speed of sound estimations are computed by both models when compared to the data in this work. In the low-pressure limit of this research, deviations range from (−0.01 to −0.06) % for the 5 % hydrogen mixture and (−0.02 to −0.12) % for the 10 % hydrogen mixture with GERG-2008 [8] as the reference, and (−0.1 to −0.25) % for the 50 % hydrogen mixture with AGA8-DC92 [10] as the reference. In the high-pressure limit of this work, differences range from (−0.06 to 0.06) % for the 5 mol-% hydrogen mixture, and from (−0.15 to 0.03) % for the 10 mol-% hydrogen mixture with GERG-2008 [8] as the reference, and from (−0.10 to 0.15) % for the 50 mol-% hydrogen mixture with AGA8-DC92 [10] as the reference.

No speed of sound data were found in the literature at the time of writing this paper for a binary methane + hydrogen mixture in the gas phase. However, the same mixtures were recently studied by Hernández-Gómez et al. [15], who reported accurate gas phase density using a single-sinker densimeter in the ranges $T$ = (240 to 350) K and $p$ = (1 up to 20) MPa. Relative deviations in density are of the same order of magnitude as the relative differences in speed of sound under the same conditions, although the trends are not similar. Relative deviations in density are nearly identical for both AGA8-DC92 [10] and GERG-2008 EoS [8] for the three mixtures, and relative deviations in density from GERG-2008 EoS [8] for the (0.50 $CH_4$ + 0.50 $H_2$) mixture present the lowest values of the three gas samples and remain well within model uncertainty.

Table 8 shows that the relative deviations between isobaric heat capacities as perfect-gas, $C_{p,m}^{pg}$, derived from speed of sound data in this work, and predictions from AGA8-DC92 [10] and GERG-2008 models [8], range from (−0.03 to 0.5) % for the (0.95 $CH_4$ + 0.05 $H_2$) mixture, from (0.2 to 1.1) % for the (0.90 $CH_4$ + 0.10 $H_2$) mixture, and from (0.7 to 2.2) % for the (0.50 $CH_4$ + 0.50 $H_2$) mixture, with an experimental expanded ($k$ = 2) uncertainty of 0.08 %. The discussion is equivalent for $C_{v,m}^{pg}$. $C_{p,m}^{pg}$ results are fitted to equation (15):

$$\frac{C_{\mathrm{p,m}}^{\mathrm{pg}}}{R} = 3.959 + 4.63\frac{\left(193.3 \cdot 10^{1} / T\right)^{2} e^{193.3 \cdot 10^{1}/T}}{\left(e^{193.3 \cdot 10^{1}/T} - 1\right)^{2}} \tag{33}$$

for the ($0.95\ CH_4 + 0.05\ H_2$) mixture and:

$$\frac{C_{\mathrm{p,m}}^{\mathrm{pg}}}{R} = 3.950 + 4.97\frac{\left(200.0 \cdot 10^{1} / T\right)^{2} e^{200.0 \cdot 10^{1}/T}}{\left(e^{200.0 \cdot 10^{1}/T} - 1\right)^{2}} \tag{34}$$

for the ($0.90\ CH_4 + 0.10\ H_2$) mixture, with the RMS of the residuals not exceeding 0.04 % and falling within the relative expanded ($k = 2$) experimental uncertainty.

EoS uncertainty in the ideal gas heat capacity for these mixtures is the sum of the uncertainty of $C_{\mathrm{p,m}}^{\mathrm{pg}}$ for pure methane, which is above 0.07 % compared to the speed of sound data of Lemming and Goodwin [17], and the uncertainty of $C_{\mathrm{p,m}}^{\mathrm{pg}}$ for pure hydrogen, which remains within 0.02 % for the temperature range used in this research [18]. Thus, the relative differences in the perfect-gas properties, which are nearly independent of the model used to assess our data, are within the combined accuracy of our experimental results and the EoS predictions only for the two lowest isotherms $T$ = (273.16 and 300) K for the mixtures of (5 and 10) mol-% of hydrogen content. In other regions, deviations increase with temperature and hydrogen content outside the uncertainty, with the models underestimating our points and yielding estimations which exceed the 0.1% accuracy stated by the AGA8-DC92 [10] and GERG-2008 EoS [8] in the heat capacity for these binary mixtures.

Second and third acoustic virial coefficients $\beta_a(T)$ and $\gamma_a(T)$ are also reported in table 8. Relative deviations of $\beta_a(T)$ according to GERG-2008 EoS [8] range from (−8 to 3) % for the ($0.95\ CH_4 + 0.05\ H_2$) mixture, from (−14 to −2) % for the ($0.90\ CH_4 + 0.10\ H_2$) mixture, and from (11 to 13) % for the ($0.50\ CH_4 + 0.50\ H_2$) mixture, with similar values, albeit twice the magnitude when compared to AGA8-DC92 EoS [10]. The differences and the uncertainty increase with temperature, and the model overpredicts our data for the mixtures with (5 and 10) mol-% of hydrogen content and underestimates our finding for the 50 mol-% of hydrogen mixture. In any case, discrepancies are not within the relative experimental expanded ($k$ = 2) uncertainty between (1 up to 8) %. The result of $\beta_a(T)$ at $T$ = 375 K for the mixtures with $x_{H_2}$ = (5 and 10) mol-% is not considered in the discussion, since its value is so close to zero that it is difficult to determine it with low uncertainty. Relative deviations of $\gamma_a(T)$ are only assessed with respect to AGA8-DC92 EoS [10] because the predictions from GERG-2008 EoS [8] that are computed using the reference thermodynamic properties software Refprop 9.1 [16] prove to be erroneous. Disagreements range from −54 % at $T$ = 325 K for the (0.50 $CH_4$ + 0.50 $H_2$) mixture up to 34 % at $T$ = 273.16 K for the (0.95 $CH_4$ + 0.05 $H_2$) mixture, with no clear behavior in terms of temperature or hydrogen content and far away from the experimental expanded ($k$ = 2) uncertainty between (2 to 17) %. This situation for the two acoustic coefficients might be explained by taking into account that for the mixtures with (5 and 10) mol-% of hydrogen the experimental ($p,\rho,T$) data in the gas phase region used to fit the interaction between methane + hydrogen, for both the AGA8-DC92 [10] and the GERG-2008 model [8], only consider points with a mole fraction above 15 mol-% hydrogen. For the mixture with 50 mol-% hydrogen, the major disagreement is more unexpected since the ($p,\rho,T$) data to regress the interaction do cover this composition. In any case, developing a binary specific function in the GERG-2008 EoS [8] for this binary system represents an overall improvement over the AGA8-DC92 model [10], with predictions that are twice as good in terms of speed of sound, according to our work.

Table 9 shows the regression parameters of the HCSW and LJ (12,6) effective intermolecular potentials obtained from the fitting process described by equations (14) to (25) combined with the representation of $C_{p,m}^{pg}$ given in equations (33) and (34) as well as the results of $\beta_a(T)$ reported in table 8. These coefficients yield the density second virial coefficients $B(T)$ and density interaction second virial coefficients $B_{12}(T)$ shown in figures 8 and 9. The average RMS of the residuals for both effective potentials are (1.3 and 2.2) % for the mixtures (0.95 $CH_4$ + 0.05 $H_2$) and (0.90 $CH_4$ + 0.10 $H_2$), respectively, and are within the relative average expanded ($k$ = 2) uncertainty $U_r(\beta_a)$ = 3.0

%. Figure 8 compares the $B(T)$ from this work with the most recent values reported by Hernández-Gómez et al. [15], which were determined with fairly low uncertainty for the same mixtures described herein. Results from each effective potential agree with each other and with Hernandez-Gómez et al. [15], and are both within the uncertainty estimated by the Monte Carlo method for our findings, which is represented by a short-dashed line for the HCSW potential and a long-dashed line for the LJ (12,6) potential. Although the predictions from both effective potentials are close when extrapolating to temperatures above the maximum experimental isotherm $T$ = 375 K used in this work, results of the HCSW potential tend towards more negative values than those of the LJ potential, and the literature [15] towards temperatures below 270 K. This behavior is also reflected when drawing comparisons with evaluations from Refprop 9.1 [16]. For this reason, agreement is better for the LJ potential, with a RMS of the relative deviations from GERG-2008 EoS [8] of (8.7 and 8.6) % for the HCSW potential and (1.5 and 2.4) % for the LJ (12,5) potential, which is within the relative expanded ($k$ = 2) uncertainty $U_r(B)$ = (27.5 and 37.8) %, for the mixtures (0.95 $CH_4$ + 0.05 $H_2$) and (0.90 $CH_4$ + 0.10 $H_2$), respectively.

**Table 9.** Regression parameters of the hard-core square well (HCSW) and Lennard-Jones (LJ (12,6)) effective potentials from the fit to the acoustic virial coefficients obtained for the ($CH_4$ + $H_2$) mixtures. $\sigma_{SW}$ is the hard-core length, $\varepsilon_{SW}$ is the well depth, and $g_{SW}$ is $\sigma_{SW}$ times the length of the square well. $\varepsilon_{LJ}$ is the depth of the potential well and $\sigma_{LJ}$ is the separation at which $U(r) = -\varepsilon_{LJ}$. RMS = root mean square.

| HCSW | | | | LJ (12,6) | | |
|---|---|---|---|---|---|---|
| $\sigma_{SW}$ / Å | $g_{SW}$ | $\varepsilon_{SW}$ / eV | RMS / % | $\sigma_{LJ}$ / Å | $\varepsilon_{LJ}$ / eV | RMS / % |
| (0.95 $CH_4$ + 0.05 $H_2$) | | | | | | |
| 27.272 | 1.165 | 0.04249 | 2.4 | 3.734 | 0.01241 | 2.1 |
| | | | (0.90 $CH_4$ + 0.10 $H_2$) | | | |
| 28.233 | 1.197 | 0.03544 | 3.2 | 3.682 | 0.01192 | 2.9 |

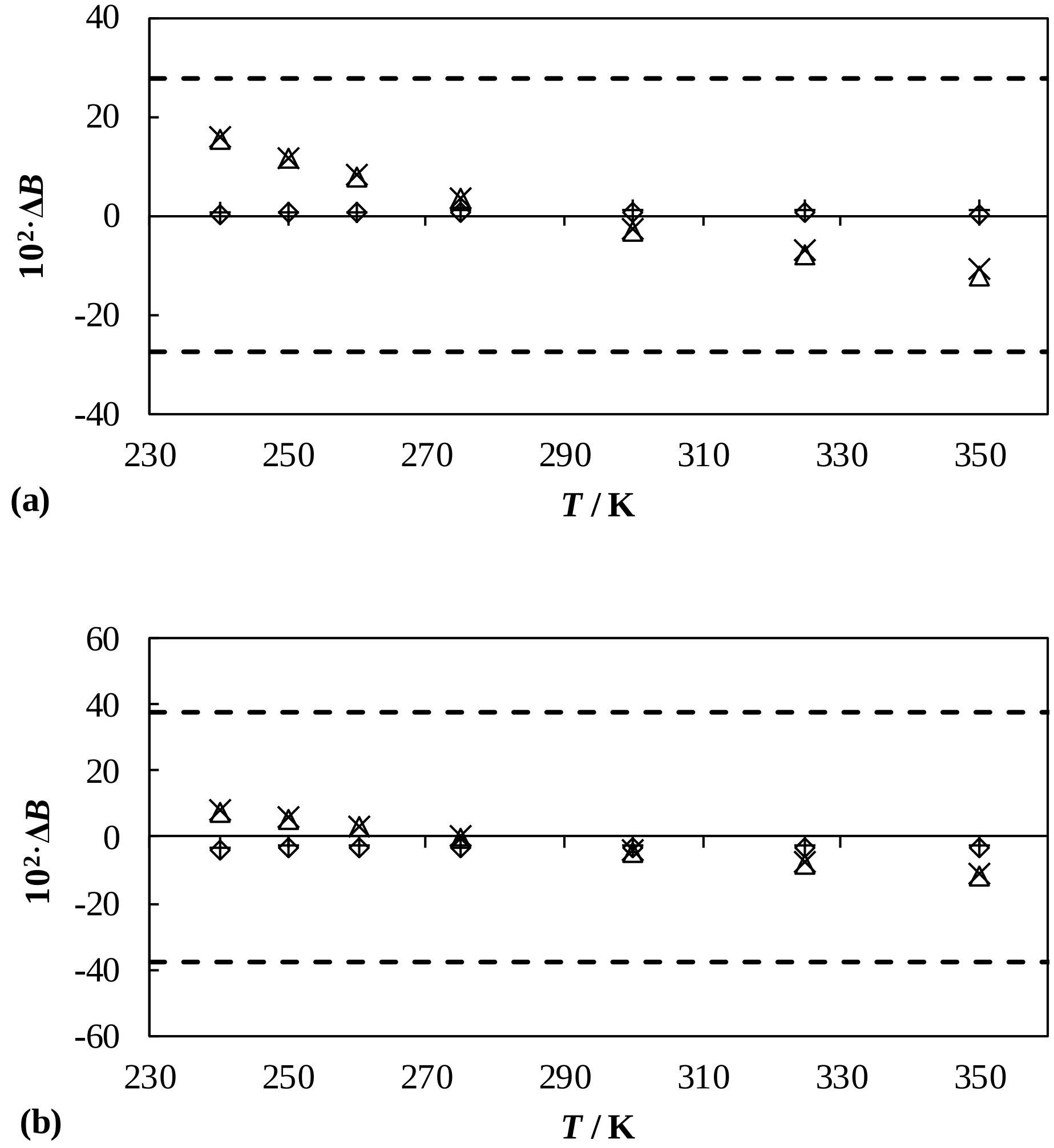

**Figure 8.** Relative deviations $\Delta B = (B_{exp} - B_{ref})/B_{ref}$ as a function of temperature for mixtures (a): (0.95 $CH_4$ + 0.05 $H_2$) and (b): (0.90 $CH_4$ + 0.10 $H_2$) from the literature values of Hernández-Gómez et al. [15] with respect to deductions from △ HCSW and ◇ LJ (12,6) potentials, and from GERG-2008 EoS [8] according to derivations from × HCSW and + LJ (12,6) potential. The short-dashed line illustrates the expanded ($k$ = 2) uncertainty of this work.

The average $B_{12}(T)$ obtained by equation (25) from the HCSW and LJ (12,6) effective potential is depicted in figure 9 as a solid line, together with the density interaction second virial coefficients from AGA8-DC92 EoS [10] (dotted line), GERG-2008 EoS [8] (dashed line), and the literature values from Mueller et al. [49], Mihara et al. [50], and Hernández-Gómez et al. [15]. The uncertainty of the coefficients derived from our speed of sound data increases after each step of the derivation process as is characteristic of the Monte Carlo procedure, and becomes the same order of magnitude as the final values. Determining the interaction coefficient $B_{12}(T)$ is highly sensitive to the mixture coefficient $B(T)$ since the density second virial coefficients $B_{11}(T)$ and $B_{22}(T)$ of pure methane and hydrogen differ enormously. $B_{11}(T)$ is negative and decreases, with a sharp dependence towards low temperature, since $B_{22}(T)$ takes positive values, decreasing with a smoother trend in the temperature range used in this study. Therefore, although all the results nearly converge at the highest temperature, for temperatures below 270 K there are numerous discrepancies. The $B_{12}(T)$ results to emerge from this research are in good agreement with the Mihara et al. data [50] for $320 < T/\mathrm{K}$, with the accurate data of Hernández-Gómez et al. [15] at $260 < T/\mathrm{K} < 325$, and show a similar trend to AGA8-DC92 [10] predictions towards low temperatures, albeit with smaller values. However, the data of Mueller et al. [49] display a different pattern, while the AGA8-DC92 [10] and GERG-2008 [8] estimations, which are almost equal at $250 < \mathrm{T/K}$, are more in line with the Hernández-Gómez et al. [15] data than with ours at temperatures above 350 K.

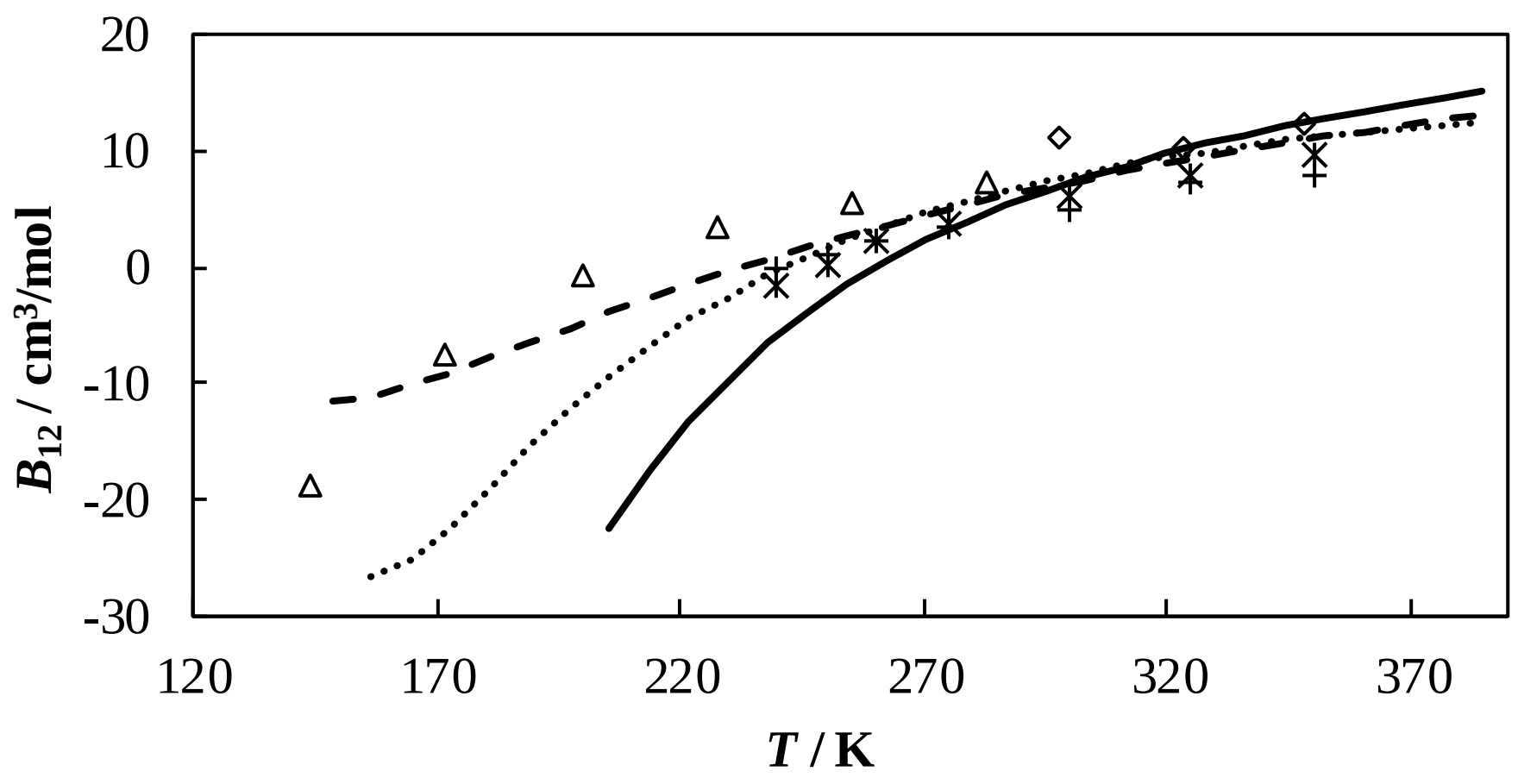

**Figure 9.** Density second interaction virial coefficient $B_{12}(T)$ as function of temperature for the ($CH_4$ + $H_2$) binary mixture: solid line from experimental values, dotted line from AGA8-DC92 EoS [10], dashed line from GERG-2008 EoS [8], + Hernández-Gómez et al. for the (0.95 $CH_4$ + 0.05 $H_2$) mixture [10], ✳ Hernández-Gómez et al. for the (0.90 CH4 + 0.10 H2) mixture [10], △ Mueller et al. [49], ◇ Mihara et al. [50].

## 7. Conclusions

In this work, new highly accurate speed of sound data for three binary mixtures (0.95 $CH_4$ + 0.05 $H_2$), (0.90 $CH_4$ + 0.10 $H_2$), and (0.50 $CH_4$ + 0.50 $H_2$) were measured using a stainless-steel spherical acoustic resonator at $T$ = (273.16, 300, 325, 350, and 375) K in the pressure range $p$ = (0.5 up to 20) MPa, with an overall relative expanded uncertainty of 220 parts in $10^6$. Data were fitted to the acoustic virial equation and perfect-gas properties, and acoustic virial coefficients were obtained for each temperature: the adiabatic coefficient $\gamma^{pg}$ with $U_r(\gamma^{pg})$ = 0.02 %, the isochoric perfect-gas heat capacity $C_{v,m}^{pg}$ with $U_r(C_{v,m}^{pg})$ = 0.08 %, the isobaric perfect-gas heat capacity $C_{p,m}^{pg}$ with $U_r(C_{p,m}^{pg})$ = 0.08 %, the second acoustic virial coefficient $\beta_a$ with $U_r(\beta_a)$ = (1 to 8) %, and the third acoustic virial coefficient $\gamma_a$ with $U_r(\gamma_a)$ = (2 to 17) %.

The new data were compared to the corresponding speed of sound values from the reference models for natural gas-like mixtures: AGA8-DC92 EoS [10] and GERG-2008 EoS [8]. Relative deviations are within experimental uncertainty only for $T$ / K < 325 and $p$ / MPa < 8 for the (5 and 10) mol-% of hydrogen mixtures, but agree well within the uncertainty of both reference EoS for the remaining points, with the exception of the (0.50 $CH_4$ + 0.50 $H_2$) mixture when compared to the GERG-2008 EoS [8]. For the two lower hydrogen content mixtures, GERG-2008 EoS [8] performs better than AGA8-DC92 EoS [10]. However, for the mixture of 50 mol-% of hydrogen, the AGA8-DC92 model [10] represents our speed of sound data better. In all instances, as the molar content of hydrogen increases the discrepancies are more temperature dependent and are of greater magnitude.

Relative deviations in the perfect-gas heat capacities $C_{p,m}^{pg}$ and $C_{v,m}^{pg}$ are consistent with the uncertainty of experimental values of 0.08% plus model uncertainty at $T$ = (273.16 and 300) K only for the (5 and 10) mol-% of hydrogen mixtures. In contrast, differences for the remaining conditions are above 0.3 %, exceeding the 0.1 % margin of uncertainty stated by the two models.

The second acoustic virial coefficients $\beta_a(T)$ deviate by over 14 %, with discrepancies that increase with temperature. The third acoustic virial coefficients $\gamma_a(T)$ evidence even greater disagreement, with no clear trend with either temperature or composition.

Additionally, speed of sound data were used to derive density second virial coefficients $B(T)$ together with the interaction ones $B_{12}(T)$ from the two mixtures with the lowest hydrogen content by applying an effective intermolecular potential fitting procedure. Although the results for $B(T)$ show relatively good agreement with literature and model results and remain within experimental uncertainty, they are not accurate enough to correctly deduce the interaction $B_{12}(T)$ coefficients for these mixtures under all the circumstances.

This research provides a detailed speed of sound study under conditions that are often encountered during industry applications for three selected binary ($CH_4$ + $H_2$) mixtures with mole fractions that are representative of $H_2$-enriched natural gas-like mixtures. This work serves as a performance test for the currently established thermodynamic reference models for this kind of gas sample, namely AGA8-DC92 [10] and GERG-2008 EoS [8], and also expands the experimental database which might be employed in the future to improve the correlation of the interaction parameters for standard equations of state.

**Acknowledgements**

This work was supported by ERDF/Ministerio de Ciencia, Innovación y Universidades – Agencia Estatal de Investigación (Project ENE2017-88474-R) and ERDF/Junta de Castilla y León (Project VA280P18).